# An algorithm for 3-SAT problems

Hiroshi Tsukimoto
Tokyo Denki University
tsukimoto@c.dendai.ac.jp

April 27, 2016

**Abstract**

This paper presents an algorithm for 3-SAT problems. First, logical formulas are transformed into elementary algebraic formulas. Second, complex trigonometric functions are assigned to the variables in the elementary algebraic formulas, and the sums of the formulas are calculated. The algorithm outputs the number of satisfying assignments. The computational complexity of the algorithm is probably polynomial.

## 1. Introduction

This paper presents an algorithm for 3-SAT problems. In the former part, logical formulas are transformed into elementary algebraic formulas. Consider a CNF formula consisting of $N$ variables and $M$ clauses.

In Section 2, the identity relaxation of a CNF formula is executed. A variable usually occurs more than once in a CNF formula. The variable in different places is replaced by different variables. As the result, new $3M$ variables are introduced. Identity($X = X$) is relaxed to identical OR(IOR). The new logical formula consists of the CNF part consisting $M$ clauses and the IOR part consisting of the logical conjunction of $N$ IORs. This logical formula is called the identity relaxed logical formula. Each variable occurs only in one clause in the CNF part, and occurs only in one IOR in the IOR part. The identity relaxation by IOR is necessary for the second identity relaxation in Section 6.

In Section 3, the identity relaxed logical formulas are expressed by elementary algebra, where true=1 and false=$-1$.

In Section 4, it is shown that the elementary algebraic expressions can satisfy the axioms of Boolean algebra. The idempotent law cannot be calculated by elementary algebra.

In Section 5, the domain of the elementary algebraic formulas is extended from $\{-1,1\}$ to $[-1,1]$. Owing to this extension, the number of satisfying assignments can be calculated from the constant term. That is, the satisfiability can be checked from the constant term. The constant term can be calculated by assigning 0s to all the variables in the expansion of the identity relaxed logical formula after the processing of the idempotent law.

In Section 6, the identity relaxation of the elementary algebraic formula is executed. The second identity relaxation is executed using inverse elements. Owing to the second identity relaxation, the idempotent law can be calculated by elementary algebra, and therefore the satisfiability can be checked by elementary algebra. However, expanding the formula is needed, whose computational complexity is exponential.

Therefore, in order to delete all the non-constant terms, another method must be developed. The latter part of this paper is devoted to the problem. Let $f$ stand for the elementary algebraic formula obtained finally in the former part. One method for deleting all the non-constant terms is assigning complex trigonometric functions to the variables in $f$ and calculating a certain sum of $f$.

In Section 7, one-variable complex trigonometric functions are assigned to the variables in $f$ as follows:

$$X_j \leftarrow \exp(ia_j x) \ (1 \leq j \leq n),$$

where $a_j$ is a real number and $n$ is the number of variables, that is $n = 3M$. A certain sum of $f$ is calculated to delete all the non-constant terms. In the case of one-variable complex trigonometric functions, the computational complexity of the sum is exponential.

In Section 8, two-variable complex trigonometric functions are assigned to the variables in $f$ as follows:
$$X_j \leftarrow \exp(i(a_j x + b_j y)).$$
A certain sum of $f$ is calculated to delete all the non-constant terms. The numerical experiments showed that it is unclear whether the computational complexity of the sum is polynomial or not.

In Section 9, four-variable complex trigonometric functions are assigned to the variables in $f$ as follows:
$$X_j \leftarrow \exp(i(a_j x + b_j y + c_j z + d_j w)).$$
A certain sum of $f$ is calculated to delete all the non-constant terms.

In Section 10, the results of the numerical experiments of four-variable complex trigonometric functions are shown. The computational complexity of the sum is probably polynomial.

The numerical experiments of four-variable complex trigonometric functions show that the algorithm may be a polynomial-time algorithm for 3-SAT problems. Regretfully, it is unproven that the algorithm is a polynomial-time algorithm for 3-SAT problems.

## 2. Identity relaxation

Consider a CNF formula consisting $M$ clauses and $N$ variables. Logical variables are denoted by capital letters, and the variables of elementary algebraic formulas are denoted by small letters. Logical formulas are denoted by capital letters, and the elementary algebraic formulas are denoted by small letters.

Usually, a variable occurs more than once in a CNF formula. Let us replace the variable in different places by different variables. Let $X, Y, Z$ and $X_i$ stand for variables. For example,
$$(X \vee Y \vee Z) \wedge (X \vee \bar{Y} \vee Z)$$
is transformed to
$$(X_1 \vee X_2 \vee X_3) \wedge (X_4 \vee X_5 \vee X_6).$$
$X, Y$ and $Z$ are called old variables, and $X_1, X_2, X_3, X_4, X_5$ and $X_6$ are called new variables. The number of new variables is $3M$.

**Definition 1.** *Identical OR(IOR)*
*If an old variable $X$ occurs in $r$ clauses $(r \leq M)$, Identical OR is defined as follows:*
$$(e_1(X_{a_1}) \wedge \cdots \wedge e_r(X_{a_r}) \vee (\overline{e_1(X_{a_1})} \wedge \cdots \wedge \overline{e_r(X_{a_r})})),$$
*where the ith $X$ is replaced by $X_{a_i} (1 \leq i \leq r)(1 \leq a_i \leq 3M)$. $e_i(X_{a_i}) = X_{a_i}$ if the ith $X$ occurs positively, that is, as $X$. $e_i(X_{a_i}) = \overline{X_{a_i}}$, if the ith $X$ occurs negatively, that is, as $\bar{X}$. The part before the logical disjunction of the above formula is called the former part, and the part after the logical disjunction is called the latter part.*

**Example 1.** The examples of IORs
$$(X \vee Y \vee Z) \wedge (X \vee \bar{Y} \vee Z) \wedge (X \vee Y \vee \bar{Z})$$
is transformed to
$$(X_1 \vee X_2 \vee X_3) \wedge (X_4 \vee X_5 \vee X_6) \wedge (X_7 \vee X_8 \vee X_9).$$
Identical OR for $X$ ($IOR(X_1, X_4, X_7)$) is
$$(X_1 \wedge X_4 \wedge X_7) \vee (\overline{X_1} \wedge \overline{X_4} \wedge \overline{X_7}).$$
Identical OR for $Y$ ($IOR(X_2, \overline{X_5}, X_8)$) is
$$(X_2 \wedge \overline{X_5} \wedge X_8) \vee (\overline{X_2} \wedge X_5 \wedge \overline{X_8}).$$
Identical OR for $Z$ ($IOR(X_3, X_6, \overline{X_9})$) is
$$(X_3 \wedge X_6 \wedge \overline{X_9}) \vee (\overline{X_3} \wedge \overline{X_6} \wedge X_9).$$

$X$ is replaced by $X_1, X_4$ and $X_7$. Identity($X = X$) is relaxed to $(X_1 \wedge X_4 \wedge X_7) \vee (\overline{X_1} \wedge \overline{X_4} \wedge \overline{X_7})$. $(X_1 \wedge X_4 \wedge X_7) \vee (\overline{X_1} \wedge \overline{X_4} \wedge \overline{X_7}))$ is true, if and only if $X_1 = T$, $X_4 = T$ and $X_7 = T$ or $X_1 = F$, $X_4 = F$ and $X_7 = F$, where $T$ stands for true, and $F$ stands for false. $Y$ is replaced by $X_2, X_5$ and $X_8$. Identity $(Y = Y)$ is relaxed to $(X_2 \wedge \overline{X_5} \wedge X_8) \vee (\overline{X_2} \wedge X_5 \wedge \overline{X_8})$. $(X_2 \wedge \overline{X_5} \wedge X_8) \vee (\overline{X_2} \wedge X_5 \wedge$ $\overline{X_8}$)) is true, if and only if $X_2 = T$, $X_5 = F$ and $X_8 = T$ or $X_2 = F$, $X_5 = T$ and $X_8 = F$. $Z$ is the same as $Y$.

**Definition 2.** *Identity relaxed logical formula*
*A logical formula consisting of the logical conjunction of the CNF formula with new variables and all the IORs is called an identity relaxed logical formula.*

**Example 2.** An example of an identity relaxed logical formula
$$(X \lor Y \lor Z) \land (X \lor \bar{Y} \lor Z) \land (X \lor Y \lor \bar{Z})$$
is transformed to
$$(X_1 \lor X_2 \lor X_3) \land (X_4 \lor X_5 \lor X_6) \land (X_7 \lor X_8 \lor X_9).$$
The logical conjunction of the above formula and IORs is as follows:
$$(X_1 \lor X_2 \lor X_3) \land (X_4 \lor X_5 \lor X_6) \land (X_7 \lor X_8 \lor X_9) \land IOR(X_1, X_4, X_7) \land IOR(X_2, \overline{X_5}, X_8)$$
$$\land IOR(X_3, X_6, \overline{X_9})$$
$$= (X_1 \lor X_2 \lor X_3) \land (X_4 \lor X_5 \lor X_6) \land (X_7 \lor X_8 \lor X_9) \land ((X_1 \land X_4 \land X_7) \lor (\overline{X_1} \land \overline{X_4} \land \overline{X_7})) \land ((X_2 \land \overline{X_5} \land X_8) \lor (\overline{X_2} \land X_5 \land \overline{X_8})) \land ((X_3 \land X_6 \land \overline{X_9}) \lor (\overline{X_3} \land \overline{X_6} \land X_9)).$$

**Theorem 1.** *The number of satisfying assignments of a CNF is equal to the number of satisfying assignments of the identity relaxed logical formula.*
*Proof.* Let $N$ and $M$ stand for the number of variables and the number of clauses, respectively. There are $2^N$ assignments for a CNF formula. There are $2^{3M}$ assignments for the identity relaxed logical formula. The $2^N$ assignments for a CNF formula are called real assignments. The assignments except real assignments for the identity relaxed logical formula are called imaginary assignments. There are $2^{3M} - 2^N$ imaginary assignments. The IOR part of the identity relaxed logical formula is true under all the real assignments, and false under all the imaginary assignments.

If a CNF formula is satisfiable under an assignment, then the CNF part of the identity relaxed logical formula is satisfiable under the assignment. The IOR part of the identity relaxed logical formula is true under all the real assignments. Therefore the identity relaxed logical formula is satisfiable under the assignment. If a CNF formula is unsatisfiable under an assignment, then the CNF part of the identity relaxed logical formula is unsatisfiable under the assignment, and therefore the identity relaxed logical formula is unsatisfiable under the assignment. Thus, it is understood that the number of satisfying assignments of a CNF is equal to the number of satisfying assignments of the identity relaxed logical formula. □

## 3. Elementary algebraic expressions

Let true=1 and false=$-1$. Let $F$ and $G$ stand for logical formulas. Table 1 shows logical conjunction, logical disjunction and negation.

**Table 1** Truth table

| $F$ | $G$ | $F \land G$ | $F \lor G$ | $\neg F$ |
|---|---|---|---|---|
| 1 | 1 | 1 | 1 | $-1$ |
| 1 | $-1$ | $-1$ | 1 | $-1$ |
| $-1$ | 1 | $-1$ | 1 | 1 |
| $-1$ | $-1$ | $-1$ | $-1$ | 1 |

**Theorem 2.** *Let $f$ and $g$ stand for elementary algebraic formulas. The elementary algebraic expressions of logical conjunction($\land$), logical disjunction($\lor$) and negation($\neg$) are as follows*:
$$F \land G \to \frac{1}{2}(f+1)(g+1) - 1,$$

$$F \vee G \to -\frac{1}{2}(f-1)(g-1) + 1,$$
$$\neg F \to -f.$$

*Proof.* It can be easily verified that the above elementary algebraic expressions satisfy Table1. □

**Theorem 3.** *The elementary algebraic expression of the IOR of r variables is as follows*:
$$\frac{1}{2^{r-1}}((e_1 x_1 + 1) \cdots (e_r x_r + 1) + (-e_1 x_1 + 1) \cdots (-e_r x_r + 1)) - 1,$$
*where* $e_i = 1 (1 \leq i \leq r)$, *if X occurs positively, and* $e_i = -1$, *if X occurs negatively.*

*Proof.* Let us assume that $e_i = 1$ for all $i$, that is, $X$ occurs positively for all $i$. Then, the above formula is transformed into
$$\frac{1}{2^{r-1}}((x_1 + 1) \cdots (x_r + 1) + (-x_1 + 1) \cdots (-x_r + 1)) - 1.$$
Only when $x_i = 1$ for all $i$ or $x_i = -1$ for all $i$, the above formula=1. The other cases can be argued in the same fashion. □

**Example 3.** An example of the elementary algebraic expression of an IOR
$$(X \vee Y \vee Z) \wedge (\bar{X} \vee Y \vee Z) \wedge (X \vee W \vee Z)$$
is transformed to
$$(X_1 \vee X_2 \vee X_3) \wedge (X_4 \vee X_5 \vee X_6) \wedge (X_7 \vee X_8 \vee X_9).$$
The IOR of $X$ is
$$(X_1 \wedge \overline{X_4} \wedge X_7) \vee (\overline{X_1} \wedge X_4 \wedge \overline{X_7})$$
The elementary algebraic expression is as follows:

$$\frac{1}{2^{3-1}}((x_1 + 1)(-x_4 + 1)(x_7 + 1) + (-x_1 + 1)(x_4 + 1)(-x_7 + 1)) - 1.$$

**Theorem 4.** *Let $F_i$ $(1 \leq i \leq r)$ stand for a logical formula, and $f_i$ stand for the elementary algebraic expression of $F_i$. The logical conjunction and disjunction of r formulas are as follows*:
$$F_1 \wedge \cdots \wedge F_r \to \frac{1}{2^{r-1}}\big((f_1 + 1) \cdots (f_r + 1)\big) - 1,$$
$$F_1 \vee \cdots \vee F_r \to \frac{1}{(-2)^{r-1}}\big((f_1 - 1) \cdots (f_r - 1)\big) + 1.$$
*Proof.* The above formulas are obtained by the repeated use of the formulas in Theorem 2. □

**Theorem 5.** *Consider an identity relaxed logical formula of N variables and M clauses*
$$C_1 \wedge \cdots \wedge C_M \wedge IOR_1 \wedge \cdots \wedge IOR_N,$$
*where $C_i$ stands for a clause. The elementary algebraic expression of the identity relaxed logical formula is as follows*:
$$\frac{1}{2^{6M-1}}\big(((x_1 - 1)(x_2 - 1)(x_3 - 1) + 8) \cdots ((x_{3M-2} - 1)(x_{3M-1} - 1)(x_{3M} - 1) + 8)$$
$$\big((e_{11} x_{11} + 1) \cdots (e_{1r_1} x_{1r_1} + 1) + (-e_{11} x_{11} + 1) \cdots (-e_{1r_1} x_{1r_1} + 1)\big)$$
$$\cdots ((e_{N1} x_{N1} + 1) \cdots (e_{Nr_N} x_{Nr_N} + 1) + (-e_{N1} x_{N1} + 1) \cdots (-e_{Nr_N} x_{Nr_N} + 1))\big) - 1,$$
*where $r_i (1 \leq i \leq N)$ stands for the number of occurrences of an old variable $X_i$ in the CNF. $x_{ij}$ $(1 \leq j \leq r_i)$ is assigned to the jth $X_i$. If $X_i$ occurs positively in the jth place, $e_{ij} = 1$, and if $X_i$ occurs negatively in the jth place, $e_{ij} = -1$.*

*Proof.*
$$C_1 \wedge \cdots \wedge C_M \wedge IOR_1 \wedge \cdots \wedge IOR_N$$
$$= (X_1 \vee X_2 \vee X_3) \wedge \cdots \wedge (X_{3M-2} \vee X_{3M-1} \vee X_{3M}) \wedge IOR_1 \wedge \cdots \wedge IOR_N.$$
Since

$$X \vee Y \vee Z \to \frac{1}{4}(x-1)(y-1)(z-1)+1,$$

the elementary algebraic expression of the above formula is as follows:

$$\frac{1}{2^{M+N-1}}\left(\frac{1}{4}(x_1-1)(x_2-1)(x_3-1)+1+1\right)\cdots\left(\frac{1}{4}(x_{3M-2}-1)(x_{3M-1}-1)(x_{3M}-1)+1+1\right)$$
$$\left(\frac{1}{2^{r_1-1}}\left((e_{11}x_{11}+1)\cdots(e_{1r_1}x_{1r_1}+1)+(-e_{11}x_{11}+1)\cdots(-e_{1r_1}x_{1r_1}+1)\right)-1+1\right)$$
$$\cdots\left(\frac{1}{2^{r_N-1}}\left((e_{N1}x_{N1}+1)\cdots(e_{Nr_N}x_{Nr_N}+1)+(-e_{N1}x_{N1}+1)\cdots(-e_{Nr_N}x_{Nr_N}+1)\right)-1+1\right)-1$$

The above formula is transformed as follows:

$$\frac{1}{2^{M+N-1}}\frac{1}{4^M}\left(((x_1-1)(x_2-1)(x_3-1)+8)\cdots((x_{3M-2}-1)(x_{3M-1}-1)(x_{3M}-1)+8)\right)$$
$$\frac{1}{2^{\sum_{i=1}^{N}(r_i-1)}}\left(((e_{11}x_{11}+1)\cdots(e_{1r_1}x_{1r_1}+1)+(-e_{11}x_{11}+1)\cdots(-e_{1r_1}x_{1r_1}+1))\right.$$
$$\left.\cdots((e_{N1}x_{N1}+1)\cdots(e_{Nr_N}x_{Nr_N}+1)+(-e_{N1}x_{N1}+1)\cdots(-e_{Nr_N}x_{Nr_N}+1))\right)-1$$
$$\to \frac{1}{2^{6M-1}}\left(((x_1-1)(x_2-1)(x_3-1)+8)\cdots((x_{3M-2}-1)(x_{3M-1}-1)(x_{3M}-1)+8)\right.$$
$$((e_{11}x_{11}+1)\cdots(e_{1r_1}x_{1r_1}+1)+(-e_{11}x_{11}+1)\cdots(-e_{1r_1}x_{1r_1}+1))$$
$$\left.\cdots((e_{N1}x_{N1}+1)\cdots(e_{Nr_N}x_{Nr_N}+1)+(-e_{N1}x_{N1}+1)\cdots(-e_{Nr_N}x_{Nr_N}+1))\right)-1. \qquad \square$$

**Example 4.** An example of an elementary algebraic expression
$$(X \vee Y \vee Z) \wedge (X \vee Y \vee \bar{Z})$$
$$\to (X_1 \vee X_2 \vee X_3) \wedge (X_4 \vee X_5 \vee X_6) \wedge IOR_1 \wedge IOR_2 \wedge IOR_3$$
$$\to (X_1 \vee X_2 \vee X_3) \wedge (X_4 \vee X_5 \vee X_6) \wedge IOR(X_1,X_4) \wedge IOR(X_2,X_5) \wedge IOR(X_3,\overline{X_6})$$

The elementary algebraic expression is as follows:

$$\frac{1}{2^{2+3-1}}\left(\frac{1}{4}(x_1-1)(x_2-1)(x_3-1)+2\right)\left(\frac{1}{4}(x_4-1)(x_5-1)(x_6-1)+2\right)$$
$$\frac{1}{2}((x_1+1)(x_4+1)+(-x_1+1)(-x_4+1))\frac{1}{2}((x_2+1)(x_5+1)+(-x_2+1)(-x_5+1))$$
$$\frac{1}{2}((x_3+1)(-x_6+1)+(-x_3+1)(x_6+1))-1$$
$$\to \frac{1}{2^{11}}((x_1-1)(x_2-1)(x_3-1)+8)((x_4-1)(x_5-1)(x_6-1)+8)$$
$$((x_1+1)(x_4+1)+(-x_1+1)(-x_4+1))((x_2+1)(x_5+1)+(-x_2+1)(-x_5+1))$$
$$((x_3+1)(-x_6+1)+(-x_3+1)(x_6+1))-1.$$

## 4. The elementary algebraic expressions and the axioms of Boolean algebra

Let $f, g$ and $h$ stand for the elementary algebraic expressions of identity relaxed logical formulas.

**Theorem 6.** $f^2 = 1$ *holds.* $f^2 = 1$ *is equivalent to the idempotent law* $(f \wedge f = f \vee f = f)$.
*Proof.* Since $f = 1$ or $f = -1$, $f^2 = 1$. $f^2 = 1 \leftrightarrow f \wedge f = f \vee f = f$ can be easily checked using Theorem 2. $\qquad \square$

$x_i^2 = 1$ holds for all $i$, because $x_i = 1$ or $x_i = -1$. $f^2 = 1$ and $x_i^2 = 1$ are the elementary algebraic expressions of the idempotent law.

The axioms of Boolean algebra are as follows[1]:
1. $f \vee g = g \vee f$, $f \wedge g = g \wedge f$,
2. $f \vee (g \vee h) = (f \vee g) \vee h$, $f \wedge (g \wedge h) = (f \wedge g) \wedge h$,
3. $(f \wedge g) \vee g = g$, $(f \vee g) \wedge g = g$,
4. $f \wedge (g \vee h) = (f \wedge g) \vee (f \wedge h)$, $f \vee (g \wedge h) = (f \vee g) \wedge (f \vee h)$,

5. $(f\wedge\bar{f})\vee g = g$, $(f\vee\bar{f})\wedge g = g$.

**Theorem 7.** *The elementary algebraic expressions satisfy the above axioms.*
*Proof.* 1. and 2. can be checked by the elementary algebraic expressions. 3. can be checked by the elementary algebraic expressions and $g^2 = 1$. 4. and 5. can be checked by the elementary algebraic expressions and $f^2 = 1$. □

**Theorem 8.** *The degrees of variables in f are less than or equal to 2.*
A remark on degree: For example, let
$$f = x_1 x_2 x_3 x_4 (x_1 x_3 + 2x_3)(= x_1^2 x_2 x_3^2 x_4 + 2x_1 x_2 x_3^2 x_4),$$
then the degree of $x_1$ is 2, and the degree of $x_2$ is 1.
*Proof.* A variable occurs only once in the CNF part, therefore the degree of the variable is 1 in the CNF part. The variable occurs in only one IOR in the IOR part. The variable occurs once in the former part of the IOR, and occurs once in the latter part of the IOR. (See Definition 1 for the former part of the IOR and the latter part of the IOR.) Since the former part of the IOR and the latter part of the IOR are added in the IOR, the degree of the variable is 1 in the IOR and in the IOR part. The CNF part is multiplied by the IOR part, and therefore the degree of the variable in $f$ is less than or equal to 2. □

**Theorem 9.** *Let us expand $f$ and apply $x_i^2 = 1$ to all the variables. Let the formula be denoted by g. The degrees of all the variables in g are less than or equal to 1.*
*Proof.* From Theorem 8, the degrees of variables in $f$ are less than or equal to 2. If $x_i^2 = 1$ is applied to all the variables in the expansion of $f$, then all the $x_i^2$s are transformed into 1s. Therefore, the degrees of all the variables in $g$ are less than or equal to 1. □

**Example 5.** An example of 2-SAT is shown. An example of 3-SAT is not shown, because there are a huge number of terms in the expansions of the examples of 3-SAT.
$$(X \vee Y) \wedge (\bar{X} \vee Y)$$
$$\rightarrow (X_1 \vee X_2) \wedge (X_3 \vee X_4) \wedge ((X_1 \wedge \overline{X_3}) \vee (\overline{X_1} \wedge X_3)) \wedge ((X_2 \wedge X_4) \vee (\overline{X_2} \wedge \overline{X_4}))$$
$$\rightarrow \frac{1}{2^3}\left(-\frac{1}{2}(x_1-1)(x_2-1)+2\right)\left(-\frac{1}{2}(x_3-1)(x_4-1)+2\right)$$
$$\frac{1}{2}((x_1+1)(-x_3+1)+(-x_1+1)(x_3+1))\frac{1}{2}((x_2+1)(x_4+1)+(-x_2+1)(-x_4+1))-1.$$
Let us expand the above formula and apply $x_1^2 \rightarrow 1, x_2^2 \rightarrow 1, x_3^2 \rightarrow 1$ and $x_4^2 \rightarrow 1$, then the following formula is obtained:
$$-\frac{3}{4}+\frac{x_2}{4}-\frac{x_1 x_3}{4}-\frac{1}{4}x_1 x_2 x_3 + \frac{x_4}{4}+\frac{x_2 x_4}{4}-\frac{1}{4}x_1 x_3 x_4 - \frac{1}{4}x_1 x_2 x_3 x_4.$$
The process is omitted. For the process, see Appendix.

## 5. The number of satisfying assignments is calculated from the constant term.

In this section, let $g$ in Theorem 9 be denoted by $f$ or $f(x_1, x_2 \cdots x_n)$, where $n = 3M$.

**Theorem 10.** *The following formula holds;*
$$f(x_1, x_2 \cdots x_n) = \frac{1}{2^n} \sum_{j=1}^{2^n} f(e_1, e_2, \cdots, e_n)(1+e_1 x_1)(1+e_2 x_2) \cdots (1+e_n x_n),$$
*where $e_i$ ($1 \leq i \leq n$) is 1 or $-1$.*
*Proof.* Let the right-hand side be denoted by $g(x_1, x_2 \cdots x_n)$. The degrees of $x_i$'s in $g(x_1, x_2 \cdots x_n)$ are 1 or 0. Therefore, $g(x_1, x_2 \cdots x_n)$ is expanded as follows:
$$g(x_1, x_2 \cdots x_n) = r_1 x_1 x_2 \cdots x_n + \cdots + r_k x_1 x_2 \cdots x_s + \cdots + r_{2^n-1} x_n + r_{2^n},$$
where $s < n$ and $1 < k < 2^n - 1$. The degrees of $x_i$'s ($1 \leq i \leq n$) in $f(x_1, x_2 \cdots x_n)$ are 1 or

0. Therefore, each term contains $x_i$ or does not contain $x_i$, and so there are $2^n$ terms in $f(x_1, x_2 \cdots x_n)$. $f(x_1, x_2 \cdots x_n)$ is expanded as follows:
$$f(x_1, x_2 \cdots x_n) = q_1 x_1 x_2 \cdots x_n + \cdots + q_k x_1 x_2 \cdots x_s + \cdots + q_{2^n-1} x_n + q_{2^n}.$$
If the values of $f(x_1, x_2 \cdots x_n)$ ($g(x_1, x_2 \cdots x_n)$) are given at $2^n$ points, then $f(x_1, x_2 \cdots x_n)(g(x_1, x_2 \cdots x_n))$ is determined uniquely. Therefore, if the values of $f(x_1, x_2 \cdots x_n)$ are equal to the values of $g(x_1, x_2 \cdots x_n)$ at $2^n$ points, then $f(x_1, x_2 \cdots x_n) = g(x_1, x_2 \cdots x_n)$.

For example, let us assign 1s to all the $x_i$'s ($1 \leq i \leq n$) in
$$f(x_1, x_2 \cdots x_n) = \frac{1}{2^n} \sum_{j=1}^{2^n} f(e_1, e_2, \cdots, e_n)(1 + e_1 x_1)(1 + e_2 x_2) \cdots (1 + e_n x_n).$$
Then, the left-hand side becomes $f(1,1 \cdots,1)$. The right-hand side becomes
$$\frac{1}{2^n} \sum_{j=1}^{2^n} f(e_1, e_2, \cdots, e_n)(1 + e_1)(1 + e_2) \cdots (1 + e_n).$$
When $e_1 = e_2 = \cdots = e_n = 1$,
$$\frac{1}{2^n} f(e_1, e_2 \cdots e_n)(1 + e_1)(1 + e_2) \cdots (1 + e_n) = f(1,1 \cdots,1).$$
Otherwise,
$$\frac{1}{2^n} f(e_1, e_2 \cdots e_n)(1 + e_1)(1 + e_2) \cdots (1 + e_n) = 0.$$
Therefore,
$$\frac{1}{2^n} \sum_{j=1}^{2^n} f(e_1, e_2, \cdots, e_n)(1 + e_1)(1 + e_2) \cdots (1 + e_n) = f(1,1 \cdots,1).$$
Thus, the left-hand side is equal to the right-hand side. Under the other assignments, the left-hand side=the right-hand side can be shown in the same manner. Consequently, $f(x_1, x_2 \cdots x_n)$ is equal to $g(x_1, x_2 \cdots x_n)$ at $2^n$ points, and so $f(x_1, x_2 \cdots x_n) = g(x_1, x_2 \cdots x_n)$. □

Theorem 10 corresponds to the expansions of Boolean functions by atoms (the minimal terms).

**Example 6.** An example of Theorem 10
$$f(x, y) = \frac{1}{4}(f(-1,-1)(1-x)(1-y) + f(-1,1)(1-x)(1+y)$$
$$+ f(1,-1)(1+x)(1-y) + f(1,1)(1+x)(1+y))$$

**Example 7.** The function of Example 5 is expanded as follows:
$$f(x_1, x_2, x_3, x_4) = -\frac{3}{4} + \frac{x_2}{4} - \frac{x_1 x_3}{4} - \frac{1}{4} x_1 x_2 x_3 + \frac{x_4}{4} + \frac{x_2 x_4}{4} - \frac{1}{4} x_1 x_3 x_4 - \frac{1}{4} x_1 x_2 x_3 x_4$$
$$= \frac{1}{16}(f(-1,-1,-1,-1)(1-x_1)(1-x_2)(1-x_3)(1-x_4)$$
$$+ f(-1,-1,-1,1)(1-x_1)(1-x_2)(1-x_3)(1+x_4)$$
$$+ f(-1,-1,1,-1)(1-x_1)(1-x_2)(1+x_3)(1-x_4)$$
$$+ f(-1,-1,1,1)(1-x_1)(1-x_2)(1+x_3)(1+x_4)$$
$$+ f(-1,1,-1,-1)(1-x_1)(1+x_2)(1-x_3)(1-x_4)$$
$$+ f(-1,1,-1,1)(1-x_1)(1+x_2)(1-x_3)(1+x_4)$$
$$+ f(-1,1,1,-1)(1-x_1)(1+x_2)(1+x_3)(1-x_4)$$
$$+ f(-1,1,1,1)(1-x_1)(1+x_2)(1+x_3)(1+x_4)$$
$$+ f(1,-1,-1,-1)(1+x_1)(1-x_2)(1-x_3)(1-x_4)$$
$$+ f(1,-1,-1,1)(1+x_1)(1-x_2)(1-x_3)(1+x_4)$$
$$+ f(1,-1,1,-1)(1+x_1)(1-x_2)(1+x_3)(1-x_4)$$
$$+ f(1,-1,1,1)(1+x_1)(1-x_2)(1+x_3)(1+x_4)$$
$$+ f(1,1,-1,-1)(1+x_1)(1+x_2)(1-x_3)(1-x_4)$$

$$
\begin{aligned}
&+f(1,1,-1,1)(1 + x_1)(1 + x_2)(1 - x_3)(1 + x_4)\\
&+f(1,1,1,-1)(1 + x_1)(1 + x_2)(1 + x_3)(1 - x_4)\\
&+f(1,1,1,1)(1 + x_1)(1 + x_2)(1 + x_3)(1 + x_4))
\end{aligned}
$$

$$
\begin{aligned}
= \frac{1}{16}(&-(1 - x_1)(1 - x_2)(1 - x_3)(1 - x_4) - (1 - x_1)(1 - x_2)(1 - x_3)(1 + x_4)\\
&-(1 - x_1)(1 - x_2)(1 + x_3)(1 - x_4) - (1 - x_1)(1 - x_2)(1 + x_3)(1 + x_4)\\
&-(1 - x_1)(1 + x_2)(1 - x_3)(1 - x_4) - (1 - x_1)(1 + x_2)(1 - x_3)(1 + x_4)\\
&-(1 - x_1)(1 + x_2)(1 + x_3)(1 - x_4) + (1 - x_1)(1 + x_2)(1 + x_3)(1 + x_4)\\
&-(1 + x_1)(1 - x_2)(1 - x_3)(1 - x_4) - (1 + x_1)(1 - x_2)(1 - x_3)(1 + x_4)\\
&-(1 + x_1)(1 - x_2)(1 + x_3)(1 - x_4) - (1 + x_1)(1 - x_2)(1 + x_3)(1 + x_4)\\
&-(1 + x_1)(1 + x_2)(1 - x_3)(1 - x_4) + (1 + x_1)(1 + x_2)(1 - x_3)(1 + x_4)\\
&-(1 + x_1)(1 + x_2)(1 + x_3)(1 - x_4) - (1 + x_1)(1 + x_2)(1 + x_3)(1 + x_4)).
\end{aligned}
$$

The domain of $f(= f(x_1, x_2 \cdots x_n))$ is extended from $\{-1,1\}^n$ to $[-1,1]^n$.

**Theorem 11.** *Let the number of satisfying assignments be $k$, the constant term of $f$ is as follows*:
$$\frac{1}{2^{3M}}(2k - 2^{3M}).$$

*Proof.* Let us assign 0s to all the $x_i$'s $(1 \leq i \leq n)$ in $f(x_1, x_2 \cdots x_n)$ in Theorem 10, then the below formula is obtained:
$$f(0,0 \cdots 0) = \frac{1}{2^n} \sum_{j=1}^{2^n} f(e_1, e_2, \cdots, e_n) \ (e_i = -1 \text{ or } 1).$$
The left-hand side is the constant term of $f$, because $f$ is a polynomial. The right-hand side is the sum of all the truth values over the whole domain. Since $n = 3M$, the number of satisfying assignments is $k$, and the number of unsatisfying assignments is $2^{3M} - k$, the right-hand side is as follows:
$$\frac{1}{2^{3M}}(k + (-1)(2^{3M} - k)) = \frac{1}{2^{3M}}(2k - 2^{3M}).$$
Thus, the theorem has been proved. □

**Theorem 12.** *If $f$ is unsatisfiable, the constant term is $-1$.*
*Proof.* If $f$ is unsatisfiable, $k = 0$ and the constant term is $-1$ from Theorem 11. □

If $f$ is a tautology, $k = 2^N$, and the constant term is $\frac{1}{2^{3M}}(2^{(N+1)} - 2^{3M})$. The constant term of a tautology is not 1, because there are $2^{3M} - 2^N$ imaginary assignments, which cannot satisfy the IOR part of $f$, and therefore cannot satisfy $f$.

**Example 8.** In the case of 2-SAT, there are two literals in one clause, therefore the constant term is not $\frac{1}{2^{3M}}(2k - 2^{3M})$ (Theorem 11), but $\frac{1}{2^{2M}}(k - (2^{2M} - k)) = \frac{1}{2^{2M}}(2k - 2^{2M})$.
Example 5 showed that
$$(X \vee Y) \wedge (\bar{X} \vee Y)$$
is transformed to
$$-\frac{3}{4} + \frac{x_2}{4} - \frac{x_1 x_3}{4} - \frac{1}{4}x_1 x_2 x_3 + \frac{x_4}{4} + \frac{x_2 x_4}{4} - \frac{1}{4}x_1 x_3 x_4 - \frac{1}{4}x_1 x_2 x_3 x_4.$$
The constant term is $-\frac{3}{4}$. The number of satisfying assignments of $(X \vee Y) \wedge (\bar{X} \vee Y)$ is 2 $((X,Y) = (-1,1), (1,1))$, and so $k = 2$. Since $M = 2$ and $k = 2$, $\frac{1}{2^{2M}}(2k - 2^{2M}) = -\frac{3}{4}$.

## 6. Identity relaxation by inverse element

The idempotent law $x_i^2 = 1$ cannot be calculated by elementary algebra. In other words, $x_i^2$ cannot be transformed to 1 by elementary algebra. Let $f$ stand for the elementary algebraic expression of an identity relaxed logical formula. Replacing $x_i$ in the IOR part by $1/x_i$ is called the identity relaxation by inverse element.

**Theorem13** *By replacing $x_i$ in the IOR part of $f$ by $\frac{1}{x_i}$ for all $i$, the idempotent law can be calculated by elementary algebra. The constant term obtained by the identity relaxation by inverse element equals the constant term obtained by the idempotent law.*

*Proof.* Recall that, in the CNF part, $x_i$ occurs only as $x_i$, that is, $x_i$ does not occur as $x_i^m (m \geq 2)$, and in the IOR part, $x_i$ occurs only as $x_i$, that is, $x_i$ does not occur as $x_i^m (m \geq 2)$. Notice that $x_i^2 = 1$ is transformed to $x_i = 1/x_i$. Let $x_i$ in the IOR part be replaced by $1/x_i$, and let $x_i$ in the CNF part be multiplied by $1/x_i$ in the IOR part, then $x_i * \frac{1}{x_i} = 1$. Thus, the idempotent law can be calculated by elementary algebra.

The constant term obtained by the idempotent law comes from the sum of the following terms in the expansion of $f$:

$$q_k \prod_{i=1}^{l} x_i^{e_i},$$

where $q_k$ is a coefficient, $1 \leq k \leq 3^n$, $l \leq n$ ($n$ is the number of variables), $e_i = 0$ or $2$ ($e_i \neq 1$). For example, $3x_1^2 x_3^2 x_4^2$ is transformed to 3, and $3x_1^2 x_3^2 x_4$ is transformed to $3 x_4$. By replacing $x_i$ in the IOR part of $f$ by $\frac{1}{x_i}$ for all $i$, $x_i^2 (= x_i \times x_i)$ is transformed to $x_i \times \frac{1}{x_i} = 1$, and therefore $q_k \prod_{i=1}^{l} x_i^{e_i}$ is transformed to $q_k \prod_{i=1}^{l} 1 = q_k$. Thus, the constant term obtained by the identity relaxation by inverse element equals the constant term obtained by the idempotent law. □

Let $g$ stand for the formula obtained by replacing $x_i$ by $1/x_i$ for all $i$ in the IOR part of $f$. Let $h$ stand for the expansion of $g$.

**Example 9.** The CNF of Example 5 is as follows:
$$(X \vee Y) \wedge (\bar{X} \vee Y).$$
The identity relaxed logical formula is as follows:
$$(X_1 \vee X_2) \wedge (X_3 \vee X_4) \wedge ((X_1 \wedge \overline{X_3}) \vee (\overline{X_1} \wedge X_3)) \wedge (X_2 \wedge X_4) \vee (\overline{X_2} \wedge \overline{X_4}).$$
The elementary algebraic expression of the identity relaxed logical formula is as follows:

$$f = \frac{1}{2^3}\left(-\frac{1}{2}(x_1-1)(x_2-1)+2\right)\left(-\frac{1}{2}(x_3-1)(x_4-1)+2\right)$$
$$\frac{1}{2}((x_1+1)(-x_3+1)+(-x_1+1)(x_3+1))\frac{1}{2}((x_2+1)(x_4+1)+(-x_2+1)(-x_4+1))-1.$$

By replacing $x_1, x_2, x_3$ and $x_4$ in the IOR part by $1/x_1, 1/x_2, 1/x_3$ and $1/x_4$, $f$ is transformed to $g$.

$$g = \frac{1}{2^3}\left(-\frac{1}{2}(x_1-1)(x_2-1)+2\right)\left(-\frac{1}{2}(x_3-1)(x_4-1)+2\right)\frac{1}{2}\left(\left(\frac{1}{x_1}+1\right)\left(-\frac{1}{x_3}+1\right)\right.$$
$$\left.+\left(-\frac{1}{x_1}+1\right)\left(\frac{1}{x_3}+1\right)\right)\frac{1}{2}\left(\left(\frac{1}{x_2}+1\right)\left(\frac{1}{x_4}+1\right)+\left(-\frac{1}{x_2}+1\right)\left(-\frac{1}{x_4}+1\right)\right)-1.$$

$g$ is expanded to $h$.

$$h = -\frac{3}{4} - \frac{1}{16x_1} + \frac{x_1}{16} + \frac{1}{8x_2} + \frac{3}{32x_1x_2} + \frac{x_1}{32x_2} + \frac{x_2}{8} - \frac{x_2}{32x_1} - \frac{3x_1x_2}{32} - \frac{1}{16x_3} - \frac{5}{16x_1x_3} - \frac{1}{32x_2x_3}$$
$$- \frac{3}{32x_1x_2x_3} + \frac{3x_2}{32x_3} - \frac{3x_2}{32x_1x_3} + \frac{x_3}{16} + \frac{x_1x_3}{16} - \frac{3x_3}{32x_2} - \frac{x_1x_3}{32x_2} + \frac{x_2x_3}{32} - \frac{1}{32}x_1x_2x_3 + \frac{1}{8x_4} - \frac{1}{32x_1x_4}$$
$$- \frac{3x_1}{32x_4} + \frac{1}{4x_2x_4} - \frac{3}{32x_1x_2x_4} + \frac{3x_1}{32x_2x_4} + \frac{3}{32x_3x_4} - \frac{3}{32x_1x_3x_4} - \frac{3}{32x_2x_3x_4} - \frac{9}{32x_1x_2x_3x_4} + \frac{x_3}{32x_4}$$

$$-\frac{x_1x_3}{32x_4}+\frac{3x_3}{32x_2x_4}+\frac{x_1x_3}{32x_2x_4}+\frac{x_4}{8}+\frac{3x_4}{32x_1}+\frac{x_1x_4}{32}+\frac{x_2x_4}{32x_1}-\frac{1}{32}x_1x_2x_4-\frac{x_4}{32x_3}-\frac{3x_4}{32x_1x_3}+\frac{x_2x_4}{32x_3}$$
$$-\frac{x_2x_4}{32x_1x_3}-\frac{3x_3x_4}{32}-\frac{1}{32}x_1x_3x_4-\frac{1}{32}x_2x_3x_4+\frac{1}{32}x_1x_2x_3x_4.$$

**Theorem 14.** *By assigning 0s to all the $x_i'$s in the numerators in $h$, and assigning $\infty$s to all the $x_i'$s in the denominators in $h$, all the non $-$ constant terms in $h$ are deleted.*
*Proof.* Assigning $\infty$s to all the $x_i's$ in the denominators in $h$ is equivalent to assigning 0s to all the $1/x_i's$ in $h$. $h$ is a polynomial of $x_i's$ and $1/x_i's$, and therefore all the non-constant terms are deleted. □

Theorem 13 and Theorem 14 say that, by elementary algebra, the constant term can be calculated and the satisfiability can be checked.

The formula obtained by the idempotent law ($x_i^2=1$) in Example 5 is as follows:
$$-\frac{3}{4}+\frac{x_2}{4}-\frac{x_1x_3}{4}-\frac{1}{4}x_1x_2x_3+\frac{x_4}{4}+\frac{x_2x_4}{4}-\frac{1}{4}x_1x_3x_4-\frac{1}{4}x_1x_2x_3x_4.$$

The constant term of $h$ in Example 9 equals the constant term of the above formula. There are more terms in $h$ in Example 9 than in the above formula. The reason is that, in $h$ in Example 9, $x_i$, which comes from the CNF part of $g$, differs from $1/x_i$, which comes from the IOR part of $g$. For example, the second term and the third term in $h$ in Example 9 is $-\frac{1}{16x_1}+\frac{x_1}{16}$, which is deleted by replacing $1/x_1$ by $x_1$. The above formula obtained by the idempotent law ($x_i^2=1$) and $h$ are equivalent as logical functions, because $\frac{1}{x_i}=x_i$ when $x_i=1$ or $x_i=-1$.

By assigning 0s to all the $x_i$'s in the numerators in $h$ which come from the CNF part of $g$ and assigning $\infty$s to all the $x_i$'s in the denominators in $h$ which come from the IOR part of $g$, the non-constant terms are deleted and the constant term is obtained. However, obtaining $h$ needs expanding $g$. The computational complexity of the expansion is exponential. It is possible to assign 0s to all the $x_i$'s in the CNF part of $g$ and assigning $\infty$s to all the $x_i$'s in the IOR part of $g$. However, the processing of the idempotent law is not executed. The non-constant terms in $g$ must be deleted without expanding $g$. The rest of this paper is devoted to this problem.

## 7. Assigning one-variable complex trigonometric functions to the variables in $f$

Hereinafter, $g$ in Section 6 is called "logical formula" and is denoted by $f$. Let $n$ stand for the number of variables, that is, $n=3M$, and let $x_j$ in $f$ be denoted by $X_j$, because $x_j$ is used for complex trigonometric functions.

In order to obtain the constant term, all the non-constant terms must be deleted. First, one-variable complex trigonometric functions are assigned to the variables in $f$. One-variable complex trigonometric functions do not work well. Second, two-variable complex trigonometric functions are assigned. It is unclear whether two-variable complex trigonometric functions work well or not. Finally, four-variable complex trigonometric functions are assigned. Four-variable complex trigonometric functions work well.

### 7.1 Assigning one-variable complex trigonometric functions
$f$ is expanded as follows:
$$f=\sum_{k=1}^{3^n-1}q_k\prod_{j=1}^{n}X_j^{e_j}+C,$$
where $C$ is the constant term, $q_k$ is a coefficient, and $e_j=-1,0$ or 1. Let us assign one-variable complex trigonometric functions to the variables in $f$ as follows:
$$X_j\leftarrow\exp(ia_jx)\,(1\leq j\leq n).$$
$a_j$'s are real numbers. All the $a_j$'s are different from each other. How to determine $a_j$'s will be

discussed later. $f$ is transformed as follows:
$$f = \sum_{k=1}^{3^n-1} q_k \prod_{j=1}^{n} (\exp(ia_j x))^{e_j} + C = \sum_{k=1}^{3^n-1} q_k \exp\left(i\left(\sum_{j=1}^{n} e_j a_j\right)x\right) + C.$$
$f$ has been transformed from a $n$-variable function to a one-variable function.

**Example 10.** An example of a non-constant term is as follows:
$$\frac{X_1}{X_2} \to \exp(i(a_1 - a_2)x).$$

### 7.2 How to delete a non-constant term
Consider a non-constant term. By ignoring the coefficient $q_k$, a non-constant term is
$$\exp\left(i\left(\sum_{j=1}^{n} e_j a_j\right)x\right),$$
where $e_j = -1, 0$ or $1$. Let
$$A_k = \sum_{j=1}^{n} e_j a_j.$$
Consider the integerization of all the $A_k$'s.

**Definition 3.** *Integerization*
*Let all the $a_j$s be multiplied by a certain integer and be rounded to non $-$ zero integers.*
*Let the integerized $a_j$ be denoted by $z(a_j)$, and let $\sum_{j=1}^{n} e_j\, z(a_j)$ be denoted by $z(A_k)$.*
*$z(\cdot)$ is called integerization.*

$$\min_{1 \le k \le 3^n-1}\{|A_k|\}\, (= \min_{1 \le k \le 3^n-1}\{|\sum_{j=1}^{n} e_j a_j|\})$$
is called the minimal frequency.
$$\max_{1 \le k \le 3^n-1}\{|A_k|\}\, (= \max_{1 \le k \le 3^n-1}\{|\sum_{j=1}^{n} e_j a_j|\})$$
is called the maximal frequency.
$$\min_{1 \le k \le 3^n-1}\{|z(A_k)|\}\, (= \min_{1 \le k \le 3^n-1}\{|\sum_{j=1}^{n} e_j\, z(a_j)|\})$$
is called the integerized minimal frequency.
$$\max_{1 \le k \le 3^n-1}\{|z(A_k)|\}\, (= \max_{1 \le k \le 3^n-1}\{|\sum_{j=1}^{n} e_j\, z(a_j)|\})$$
is called the integerized maximal frequency.

After the integerization, by ignoring the coefficient $q_k$, a non-constant term is
$$\exp(iz(A_k)x)\, (= \exp(i(\sum_{j=1}^{n} e_j\, z(a_j))x)).$$
Let us assign $\frac{2\pi}{l}m\,(1 \le m \le l)$ to $x$, and calculate the following sum:
$$S = \sum_{m=1}^{l} \exp\left(iz(A_k)\frac{2\pi}{l}m\right).$$

**Theorem 15.** *Let $t$ be a non $-$ zero integer. Let*

$$S = \sum_{m=1}^{l} \exp\left(i\frac{2\pi t}{l}m\right)\left(= \sum_{m=1}^{l} \cos\left(\frac{2\pi t}{l}m\right) + i \sum_{m=1}^{l} \sin\left(\frac{2\pi t}{l}m\right)\right).$$

If $|t| < l$, $S = 0$.

*Proof.* The following formulas hold [2]:

$$\sum_{m=1}^{l} \cos(mx) = \frac{\cos\left(\frac{(l+1)x}{2}\right) \sin\left(\frac{lx}{2}\right)}{\sin\left(\frac{x}{2}\right)},$$

$$\sum_{m=1}^{l} \sin(mx) = \frac{\sin\left(\frac{(l+1)x}{2}\right) \sin\left(\frac{lx}{2}\right)}{\sin\left(\frac{x}{2}\right)}.$$

Let us assign $\frac{2\pi t}{l}$ to $x$ in the former formula (the sum of cosine), then the following formula is obtained:

$$\sum_{m=1}^{l} \cos\left(m\frac{2\pi t}{l}\right) = \frac{\cos((l+1)\frac{\pi t}{l}) \sin(\pi t)}{\sin(\frac{\pi t}{l})}.$$

Since $t$ is a non-zero integer, $\sin(\pi t)=0$. Therefore, if $\sin(\frac{\pi t}{l}) \neq 0$, the right-hand side is 0. Since $1 \leq |t| < l$, $\sin(\frac{\pi t}{l}) \neq 0$. Therefore the right-hand side is 0. Thus,

$$\sum_{m=1}^{l} \cos\left(\frac{2\pi t}{l}m\right) = 0.$$

$$\sum_{m=1}^{l} \sin\left(\frac{2\pi t}{l}m\right) = 0$$

can be proved in the same manner. □

**Theorem16.** *If*

$$1 \leq |z(A_k)|\left(= \left|\sum_{j=1}^{n} e_j z(a_j)\right|\right) < l \ (e_j = -1, 0 \text{ or } 1)$$

*for all* $k$ $(1 \leq k \leq 3^n - 1)$,

$$\sum_{m=1}^{l} f = \sum_{m=1}^{l} \left(\sum_{k=1}^{3^n - 1} q_k \exp\left(iz(A_k)\frac{2\pi}{l}m\right) + C\right) = lC.$$

*Proof.* If

$$1 \leq |z(A_k)|\left(= \left|\sum_{j=1}^{n} e_j z(a_j)\right|\right) < l$$

for all $k$ $(1 \leq k \leq 3^n - 1)$, the sums of all the non-constant terms are 0 from Theorem 15. Therefore,

$$\sum_{m=1}^{l} f = \sum_{m=1}^{l} \sum_{k=1}^{3^n - 1} q_k \exp\left(iz(A_k)\frac{2\pi}{l}m\right) + \sum_{m=1}^{l} C = \sum_{m=1}^{l} C = lC.$$ □

There is another method to delete all the non-constant terms. The method is using

$$\int_{-\infty}^{\infty} e^{-b^2 x^2} f dx.$$

The real part of a non-constant term decreases in exponential order, because

$$\int_{-\infty}^{\infty} e^{-b^2 x^2} \cos cx \, dx = \frac{\sqrt{\pi}}{b} e^{-\frac{c^2}{4b^2}}.$$

The above integration using the trapezoidal rule converges exponentially, and therefore the computational complexity is almost the same as that of Theorem15. However, this paper does not explain this method, because it is more complicated than Theorem 15.

### 7.3 The minimal frequency and the maximal frequency

Let $a_j = \sin(n^2 + j)$. (The details will be explained in 8.2). Then, the maximal frequency is less than or equal to $n$. The author obtained the minimal frequencies by calculating all the frequencies. Table 2 shows the minimal frequencies. $m.f$ in Table 2 stands for the minimal frequency. Figure 1 shows the logarithms of the minimal frequencies. As Figure 1 shows, the minimal frequency decreases in exponential order. As the minimal frequency decreases in exponential order, $\min_{1 \le k \le 3^n - 1}\{|z(A_k)|\} / \min_{1 \le k \le 3^n - 1}\{|A_k|\}$ is of exponential order. Therefore, the integerized maximal frequency ($\max_{1 \le k \le 3^n - 1}\{|z(A_k)|\}$) is of exponential order. In order to delete all the non-constant terms using Theorem 16, $l$ is greater than the integerized maximal frequency, and so $l$ is of exponential order. Therefore, the computational complexity of the sum in Theorem 16 is exponential.

**Table 2** The minimal frequencies (one-variable complex trigonometric functions)

| $n$ | $m.f$ | $n$ | $m.f$ | $n$ | $m.f$ | $n$ | $m.f$ |
|---|---|---|---|---|---|---|---|
| 2 | $2.79 \times 10^{-1}$ | 7 | $1.06 \times 10^{-3}$ | 12 | $1.26 \times 10^{-5}$ | 17 | $9.02 \times 10^{-9}$ |
| 3 | $7.45 \times 10^{-3}$ | 8 | $1.89 \times 10^{-4}$ | 13 | $6.02 \times 10^{-7}$ | 18 | $1.84 \times 10^{-8}$ |
| 4 | $1.21 \times 10^{-2}$ | 9 | $6.72 \times 10^{-5}$ | 14 | $1.16 \times 10^{-6}$ | 19 | $8.31 \times 10^{-9}$ |
| 5 | $2.18 \times 10^{-2}$ | 10 | $8.57 \times 10^{-5}$ | 15 | $7.68 \times 10^{-7}$ | 20 | $2.24 \times 10^{-10}$ |
| 6 | $9.77 \times 10^{-3}$ | 11 | $2.45 \times 10^{-7}$ | 16 | $3.76 \times 10^{-7}$ | | |

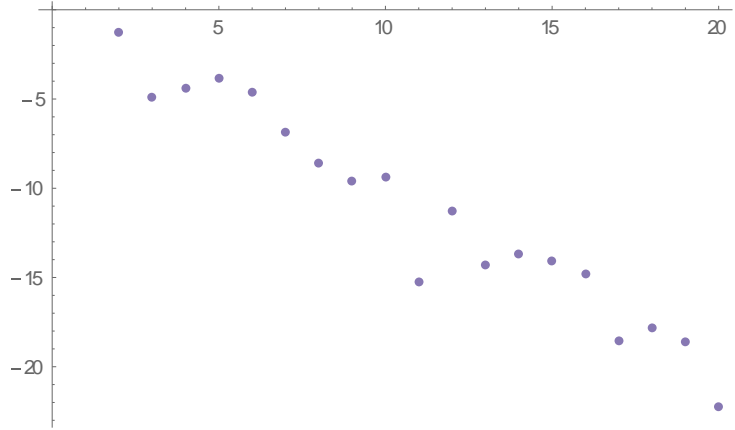

**Figure 1** The minimal frequencies (logarithm)

## 8. Assigning two-variable complex trigonometric functions to the variables in $f$

### 8.1 On the sum of two-variable complex trigonometric functions

Let us assign two-variable complex trigonometric functions to the variables in $f$ as follows:
$$X_j \leftarrow \exp(i(a_j x + b_j y)).$$
By ignoring the coefficient $q_k$, a non-constant term is as follows:
$$\exp\left(i \sum_{j=1}^{n} e_j (a_j x + b_j y)\right) = \exp\left(i \sum_{j=1}^{n} e_j a_j x\right) \exp\left(i \sum_{j=1}^{n} e_j b_j y\right) = \exp(iA_k x) \exp(iB_k y),$$
where $e_j = -1, 0$ or $1$,
$$A_k = \sum_{j=1}^{n} e_j a_j \quad \text{and} \quad B_k = \sum_{j=1}^{n} e_j b_j \quad (1 \le k \le 3^n - 1).$$

$\min_{1 \leq k \leq 3^n-1}\{\max\{|A_k|, |B_k|\}\}$ is called the minimal maximum frequency.

$\max_{1 \leq k \leq 3^n-1}\{\max\{|A_k|, |B_k|\}\}$ is called the maximal maximum frequency.

$\min_{1 \leq k \leq 3^n-1}\{\max\{|z'(A_k)|, |z'(B_k)|\}\}$ is called the intergerized minimal maximum frequency.

$\max_{1 \leq k \leq 3^n-1}\{\max\{|z'(A_k)|, |z'(B_k)|\}\}$ is called the intergerized maximal maximum frequency.

$z'(\cdot)$ stands for the integerization by which only frequencies whose absolute values are greater than or equal to the minimal maximum frequency become non-zero integers.

**Theorem 17.** Let us assign $\frac{2\pi}{l} m_1 (1 \leq m_1 \leq l)$ to x and assign $\frac{2\pi}{l} m_2 (1 \leq m_2 \leq l)$ to y. Let $\max_{1 \leq k \leq 3^n-1}\{\max\{|z'(A_k)|, |z'(B_k)|\}\} < l$.

$$S = \sum_{m_1=1}^{l} \sum_{m_2=1}^{l} \exp\left(i \frac{2\pi z'(A_k)}{l} m_1\right) \exp\left(i \frac{2\pi z'(B_k)}{l} m_2\right) = 0 \quad \text{for all } k, \quad \text{and}$$

$$S_f = \sum_{m_1=1}^{l} \sum_{m_2=1}^{l} f = l^2 C.$$

*Proof.*

$$S = \left(\sum_{m_1=1}^{l} \exp\left(i \frac{2\pi z'(A_k)}{l} m_1\right)\right)\left(\sum_{m_2=1}^{l} \exp\left(i \frac{2\pi z'(B_k)}{l} m_2\right)\right) = S_1 S_2,$$

where $S_1 = \sum_{m_1=1}^{l} \exp\left(i \frac{2\pi z'(A_k)}{l} m_1\right)$ and $S_2 = \sum_{m_2=1}^{l} \exp\left(i \frac{2\pi z'(B_k)}{l} m_2\right)$.

Since $S = S_1 S_2$, if $S_1 = 0$ or $S_2 = 0$, $S = 0$. Therefore, if at least either $z'(A_k)$ or $z'(B_k)$ is a non-zero integer (in other words, even if either one of $z'(A_k)$ or $z'(B_k)$ is zero), $S = 0$. As $z'(\cdot)$ is the integerization by which only frequencies whose absolute values are greater than or equal to the minimal maximum frequency ( $\min_{1 \leq k \leq 3^n-1}\{\max\{|A_k|, |B_k|\}\}$) become non-zero integers, all the $\max\{|z'(A_k)|, |z'(B_k)|\}$'s are non-zero integers. Therefore, $S = 0$ for all $k$. Since the sums of all the non-constant terms are deleted,

$$S_f = \sum_{m_1=1}^{l} \sum_{m_2=1}^{l} f = \sum_{m_1=1}^{l} \sum_{m_2=1}^{l} C = l^2 C. \qquad \square$$

Even if $\min_{1 \leq k \leq 3^n-1}\{|A_k|\}$ and $\min_{1 \leq k \leq 3^n-1}\{|B_k|\}$ are small, the minimal maximum frequency may not be small. If the minimal maximum frequency decreases in polynomial order, the computational complexity of $S_f$ in Theorem 17 is polynomial.

### 8.2 Frequencies are determined by trigonometric functions.

Table 3 An example of the minimal maximum frequencies

| k | $|A_k|$ | $|B_k|$ | $\max\{|A_k|, |B_k|\}$ |
|---|---------|---------|------------------------|
| 1 | 0.00103 | 0.528 | 0.528 |
| … | … | … | … |
| 1000 | 0.215 | 0.000385 | 0.215 |
| … | … | … | … |
| 2354 | 0.0000211 | 0.102 | 0.102 |
| … | … | …. | …. |
| $3^n - 1$ | 0.560 | 0.0719 | 0.560 |
| minimum | 0.0000211 | 0.000385 | 0.102 |

Table 3 shows an example of the minimal maximum frequencies. $\min_{1\le k\le 3^n-1}\{|A_k|\}$ is 0.0000211, and $\min_{1\le k\le 3^n-1}\{|B_k|\}$ is 0.000385. However, $\min_{1\le k\le 3^n-1}\{\max\{|A_k|,|B_k|\}\}$ is 0.102. As this example shows, it may be possible that the minimal maximum frequency is not small.

Let us discuss how to determine $a_j$ and $b_j$. For example, let
$$a_j = \sin t_j \text{ and } b_j = \sin(t_j + \frac{\pi}{2}),$$
where $t_j$ is a real number, then it is impossible that both of $|a_j|$ and $|b_j|$ are small at the same time. Consider $\sin jt$ $(1 \le j \le n)$, that is,
$$\sin t, \cdots, \sin jt, \cdots, \sin nt.$$
Let us assign $p$ and $p + \frac{\pi}{2}$ to $t$, then
$$\sin p, \cdots, \sin jp, \cdots, \sin np, \quad \text{and}$$
$$\sin(p + \frac{\pi}{2}), \cdots, \sin j\left(p + \frac{\pi}{2}\right), \cdots, \sin n(p + \frac{\pi}{2})$$
are obtained. If
$$a_j = \sin jp \quad \text{and} \quad b_j = \sin j(p + \frac{\pi}{2}),$$
it is possible that both of $|a_j|$ and $|b_j|$ are small at the same time for some $j$. Because the difference between the biggest frequency ($n$) and the smallest frequency (1) is large. It is necessary for all the frequencies to be similar. For example, let us replace $\sin jt$ by $\sin(n^2 + j)t$ $(1 \le j \le n)$, that is,
$$\sin(n^2 + 1)t, \cdots, \sin(n^2 + j)t, \cdots, \sin(n^2 + n)t.$$
Let
$$a_j = \sin(n^2 + j)p$$
and
$$b_j = \sin(n^2 + j)(p + \frac{\pi}{2}\frac{1}{(n^2 + q)}) \quad (1 \le q \le n),$$
then all the frequencies are similar, and therefore it is impossible that both of $|a_j|$ and $|b_j|$ are small at the same time for all $j$. Let us adopt $q = 1$, then
$$\frac{\pi}{2}\frac{1}{(n^2 + q)} = \frac{\pi}{2}\frac{1}{(n^2 + 1)}.$$
It can be expected that both of $|A_k|(= |\sum_{j=1}^n e_j a_j|)$ and $|B_k|(= |\sum_{j=1}^n e_j b_j|)$ are not small at the same time for all $k$.

### 8.3 A numerical experiment of the minimal maximum frequency.
Based on the above discussion, let us assign two-variable complex trigonometric functions to the variables in $f$ as follows:
$$X_j \leftarrow \exp(i(a_j x + b_j y)),$$
where
$$a_j = \sin((n^2 + j)p) \quad \text{and}$$
$$b_j = \sin((n^2 + j)\left(p + \frac{\pi}{2}\frac{1}{n^2 + 1}\right)).$$
By ignoring the coefficient $q_k$, a non-constant term is as follows:
$$\exp\left(i\sum_{j=1}^n e_j (a_j x + b_j y)\right) = \exp\left(i\sum_{j=1}^n e_j a_j x + i\sum_{j=1}^n e_j b_j y\right)$$
$$= \exp\left(i\sum_{j=1}^n e_j (\sin((n^2 + j)p)) x + i\sum_{j=1}^n e_j(\sin((n^2 + j)\left(p + \frac{\pi}{2}\frac{1}{n^2 + 1}\right)))y\right)$$
$$= \exp(iA_k x + iB_k y),$$
where

$$A_k = \sum_{j=1}^{n} e_j \sin((n^2+j)p) \quad \text{and}$$

$$B_k = \sum_{j=1}^{n} e_j \sin\left((n^2+j)\left(p + \frac{\pi}{2}\frac{1}{n^2+1}\right)\right),$$

where $e_j = -1, 0$ or $1$. Let $p = 1$. By calculating all the $\max\{|A_k|, |B_k|\}$'s, the minimal maximum frequencies are obtained, which are shown in Table 4. $m.m.f$ in Table 4 stands for the minimal maximum frequency. Figure 2 shows the logarithms of the minimal maximum frequencies.

**Table 4**  The minimal maximum frequencies (two-variable complex trigonometric functions)

| $n$ | $m.m.f$ | $n$ | $m.m.f$ | $n$ | $m.m.f$ | $n$ | $m.m.f$ |
|---|---|---|---|---|---|---|---|
| 2 | 0.716 | 8 | 0.00428 | 14 | 0.00185 | 20 | 0.0000477 |
| 3 | 0.0806 | 9 | 0.00390 | 15 | 0.000667 | 21 | 0.00000597 |
| 4 | 0.0605 | 10 | 0.00502 | 16 | 0.000630 | 22 | 0.0000247 |
| 5 | 0.0227 | 11 | 0.00591 | 17 | 0.000327 | 23 | 0.0000189 |
| 6 | 0.0111 | 12 | 0.00308 | 18 | 0.0000769 | 24 | 0.0000107 |
| 7 | 0.00126 | 13 | 0.00130 | 19 | 0.0000564 | 25 | 0.00000850 |

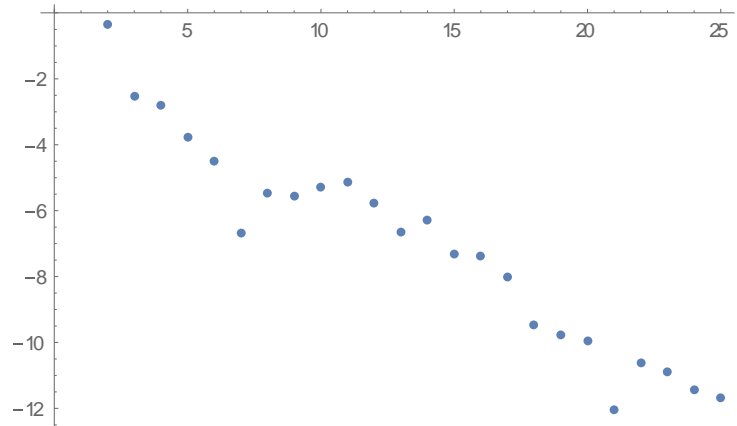

**Figure 2** The minimal maximum frequencies (logarithm)
(two-variable complex trigonometric functions)

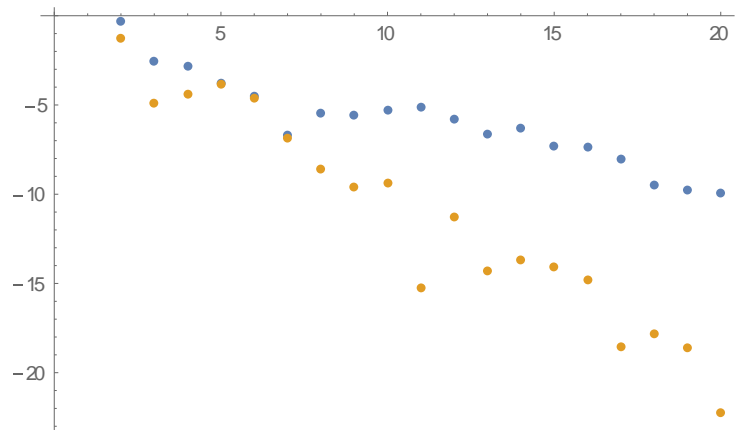

**Figure 3** The minimal frequencies of one-variable complex trigonometric functions (logarithm) and the minimal maximum frequencies of two-variable complex trigonometric functions (logarithm)

In Figure 3, yellow dots denote the logarithms of the minimal frequencies of one-variable complex trigonometric functions and blue dots denote the logarithms of the minimal maximum frequencies of two-variable complex trigonometric functions. As Figure 3 shows, the minimal maximum frequencies of two-variable complex trigonometric functions are better than the minimal frequencies of one-variable complex trigonometric functions. Figure 4 shows the approximation by a logarithm function, Figure 5 shows the approximation by an irrational function, and Figure 6 shows the approximation by a linear function. It is unclear which is the best approximation.

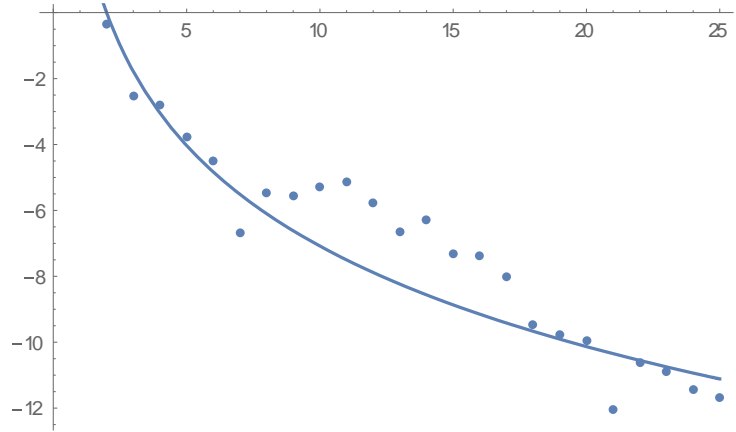

**Figure 4** $-4.40\log(0.5n)$

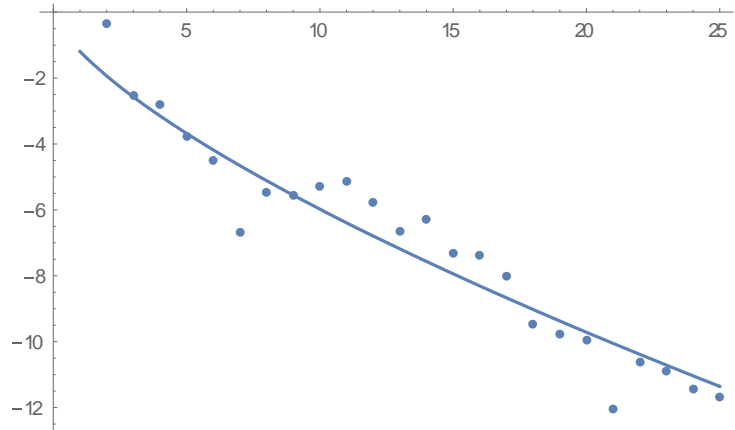

**Figure 5** $-1.19n^{0.7}$

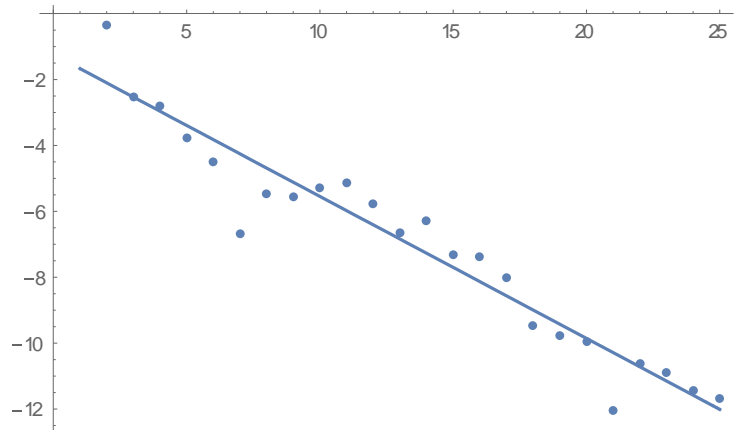

**Figure 6** $\quad -1.24 - 0.43n$

Let's investigate the result of $n = 6$. $\max\{|A_k|, |B_k|\}$ is minimized at $e_1 = -1, e_2 = 1, e_3 =$

$-1, e_4 = -1, e_5 = 1$ and $e_6 = -1$. With these $e_i$s,

$$\sum_{j=1}^{6} e_j (\sin(6^2 + j)t) = -\sin37t + \sin38t - \sin39t - \sin40t + \sin41t - \sin42t.$$

Let this function be denoted by $g(t)$, then

$$A_k = g(1) \cong -0.0111,$$
$$B_k = g\left(1 + \frac{\pi}{2}\frac{1}{6^2 + 1}\right) \cong g(1.04245) \cong 0.0000385, \text{ and}$$
$$\max\{|A_k|, |B_k|\} = 0.0111.$$

Figure 7 shows $g(t)$. As Figure 7 shows, $g(t)$ is constricted near $t = 1$ and $t = 1.04245$. The author could not delete the constrictions like this. "two-variable" in "two-variable complex trigonometric functions" corresponds to two points in Figure 7. One method for improving the results is increasing the points (=variables). Next, let us assign four-variable complex trigonometric functions.

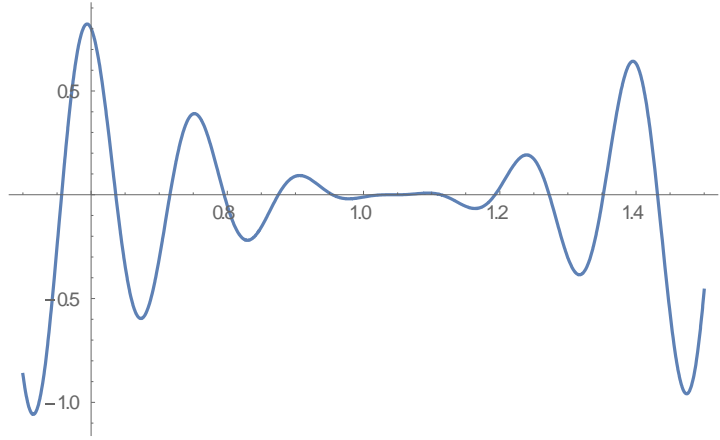

**Figure 7**  $g(t)$

## 9. Assigning four-variable complex trigonometric functions to the variables in $f$

### 9.1 Assigning four-variable complex trigonometric functions

Let us assign four-variable complex trigonometric functions to the variables in $f$ as follows:
$$X_j \leftarrow \exp(i(a_j x + b_j y + c_j z + d_j w)),$$
where
$$a_j = \sin((u + j)p),$$
$$b_j = \sin((u + j)(p + h)),$$
$$c_j = \sin((u + j)(p + v)), \text{ and}$$
$$d_j = \sin((u + j)(p + v + h)).$$

Let $g(t) = \sin((u + j)t)$, then
$$a_j = g(p), \quad b_j = g(p + h), \quad c_j = g(p + v), \quad \text{and} \quad d_j = g(p + v + h).$$

$u$ is a kind of bias as explained in 8.2. $p$ is the point for $a_j$. $v$ is the distance between the point for $c_j$ and the point for $a_j$. If $v = \frac{2\pi}{n^2+1}$, $v$ is almost one cycle of $\sin((n^2 + j)t)$. $h$, which was explained in 8.2, is the distance for preventing that the both of $|a_j|$ and $|b_j|$ (or $|c_j|$ and $|d_j|$) are small at the same time.

Let
$$u = n^2, p = 1, v = \frac{3\pi}{n^2 + 1} \quad \text{and} \quad h = \frac{\pi}{2}\frac{1}{n^2 + 1}.$$

By ignoring the coefficient $q_k$, a non-constant term is as follows:
$$\exp(iA_k x + iB_k y + iC_k z + iD_k w),$$

where

$$A_k = \sum_{j=1}^{n} e_j \sin((n^2+j)),$$

$$B_k = \sum_{j=1}^{n} e_j \sin((n^2+j)(1+\frac{\pi}{2}\frac{1}{n^2+1})),$$

$$C_k = \sum_{j=1}^{n} e_j \sin((n^2+j)(1+\frac{3\pi}{n^2+1})), \quad \text{and}$$

$$D_k = \sum_{j=1}^{n} e_j \sin\left((n^2+j)\left(1+\frac{3\pi}{n^2+1}+\frac{\pi}{2}\frac{1}{n^2+1}\right)\right),$$

where $e_j = -1, 0$ or $1$.

$\min_{1 \leq k \leq 3^n-1}\{\max\{|A_k|,|B_k|,|C_k|,|D_k|\}\}$ is called the minimal maximum frequency.

$\max_{1 \leq k \leq 3^n-1}\{\max\{|A_k|,|B_k|,|C_k|,|D_k|\}\}$ is called the maximal maximum frequency.

$\min_{1 \leq k \leq 3^n-1}\{\max\{|z'(A_k)|,|z'(B_k)|,|z'(C_k)|,|z'(D_k)|\}\}$ is called the intergerized minimal maximum frequency.

$\max_{1 \leq k \leq 3^n-1}\{\max\{|z'(A_k)|,|z'(B_k)|,|z'(C_k)|,|z'(D_k)|\}\}$ is called the intergerized maximal maximum frequency.

$z'(\cdot)$ stands for the integerization by which only frequencies whose absolute values are greater than or equal to $\min_{1 \leq k \leq 3^n-1}\{\max\{|A_k|,|B_k|,|C_k|,|D_k|\}\}$ become non-zero integers. $S$ is as follows:

$$S = \sum_{m_1=1}^{l}\sum_{m_2=1}^{l}\sum_{m_3=1}^{l}\sum_{m_4=1}^{l} \exp\left(i\frac{2\pi z'(A_k)}{l}m_1\right)\exp\left(i\frac{2\pi z'(B_k)}{l}m_2\right)\exp\left(i\frac{2\pi z'(C_k)}{l}m_3\right)\exp\left(i\frac{2\pi z'(D_k)}{l}m_4\right)$$

$$= (\sum_{m_1=1}^{l} \exp\left(i\frac{2\pi z'(A_k)}{l}m_1\right))(\sum_{m_2=1}^{l} \exp\left(i\frac{2\pi z'(B_k)}{l}m_2\right))(\sum_{m_3=1}^{l} \exp\left(i\frac{2\pi z'(C_k)}{l}m_3\right))(\sum_{m_4=1}^{l} \exp\left(i\frac{2\pi z'(D_k)}{l}m_4\right))$$

$=S_1 S_2 S_3 S_4$, where

$$S_1 = \sum_{m_1=1}^{l} \exp\left(i\frac{2\pi z'(A_k)}{l}m_1\right), S_2 = \sum_{m_2=1}^{l} \exp\left(i\frac{2\pi z'(B_k)}{l}m_2\right),$$

$$S_3 = \sum_{m_3=1}^{l} \exp\left(i\frac{2\pi z'(C_k)}{l}m_3\right) \text{ and } S_4 = \sum_{m_4=1}^{l} \exp\left(i\frac{2\pi z'(D_k)}{l}m_4\right).$$

**Theorem 18.** *If $l$ is greater than the integerized maximal maximum frequency, $S = 0$ for all $k$, and*

$$S_f = \sum_{m_1=1}^{l}\sum_{m_2=1}^{l}\sum_{m_3=1}^{l}\sum_{m_4=1}^{l} f = l^4 C,$$

where $C$ is the constant term of $f$.

*Proof.* Theorem 18 can be proved in the same manner as Theorem 17. □

### 9.2 The summary of the algorithm
1. Identity relaxation by IOR
2. Elementary algebraic expressions
3. Identity relaxation by inverse element
4. Assigning four-variable complex trigonometric functions

5. The calculation of $z'(a_j), z'(b_j), z'(c_j)$ and $z'(d_j)$
6. The calculation of the integerized maximal maximum frequency and $l$
7. The calculation of $S_f$

If $\frac{S_f}{l^4} = -1$, then the CNF formula is unsatisfiable. Otherwise it is satisfiable.

## 9.3 An example of the algorithm

**Example 11.** By identity relaxation by IOR, elementary algebraic expressions, and identity relaxation by inverse element,
$$(X \vee Y \vee Z) \wedge (X \vee Y \vee \bar{Z})$$
is transformed to
$$\frac{1}{2^{11}}\big((X_1-1)(X_2-1)(X_3-1)+8\big)\big((X_4-1)(X_5-1)(X_6-1)+8\big)$$
$$\left(\left(\frac{1}{X_1}+1\right)\left(\frac{1}{X_4}+1\right)+\left(-\frac{1}{X_1}+1\right)\left(-\frac{1}{X_4}+1\right)\right)\left(\left(\frac{1}{X_2}+1\right)\left(\frac{1}{X_5}+1\right)+\left(-\frac{1}{X_2}+1\right)\left(-\frac{1}{X_5}+1\right)\right)$$
$$\left(\left(\frac{1}{X_3}+1\right)\left(-\frac{1}{X_6}+1\right)+\left(-\frac{1}{X_3}+1\right)\left(\frac{1}{X_6}+1\right)\right) - 1.$$

For the transformation, see Theorem 5 and Theorem 13. Let us assign four-variable complex trigonometric functions. Let

$$u = n^2, p = 1, v = \frac{3\pi}{n^2+1} \text{ and } h = \frac{\pi}{2}\frac{1}{n^2+1}.$$
$$X_j \leftarrow \exp(i(a_j x + b_j y + c_j z + d_j w)) \ (1 \leq j \leq 6), \text{ where}$$
$$a_j = \sin(n^2+j),$$
$$b_j = \sin((n^2+j)\left(1+\frac{\pi}{2(n^2+1)}\right)),$$
$$c_j = \sin((n^2+j)\left(1+\frac{3\pi}{n^2+1}\right)), \text{ and}$$
$$d_j = \sin((n^2+j)\left(1+\frac{3\pi}{n^2+1}+\frac{\pi}{2(n^2+1)}\right)).$$

By calculating all the $\max\{|A_k|,|B_k|,|C_k|,|D_k|\}$'s, the minimal maximum frequency $\min_{1\leq k\leq 3^n-1}\{\max\{|A_k|,|B_k|,|C_k|,|D_k|\}\}$ is obtained. The value is 0.278. In order to integerize 0.278, 4 is an appropriate number, because $4 \times 0.278 > 1$. However, during the additions or subtractions of frequencies, the phenomena like cancellations of digits may occur, and therefore let 4 be multiplied by $n$. (Details will be explained in Section 11.) As $n = 6$, $4 \times 6 = 24$. Let us round 24 to 20. As the result, let $a_j, b_j, c_j$ and $d_j$ be multiplied by 20, and be rounded up to integers.

**Table 5**   $a_j, b_j, c_j, d_j, z'(a_j), z'(b_j), z'(c_j)$ and $z'(d)$.

| $j$ | 1 | 2 | 3 | 4 | 5 | 6 |
|---|---|---|---|---|---|---|
| $a_j$ | $-0.644$ | 0.296 | 0.964 | 0.745 | $-0.159$ | $-0.917$ |
| $z'(a_j)$ | $-13$ | 6 | 20 | 15 | $-4$ | $-19$ |
| $b_j$ | 0.765 | 0.942 | 0.184 | $-0.756$ | $-0.946$ | $-0.198$ |
| $z'(b_j)$ | 16 | 19 | 4 | $-16$ | $-19$ | $-4$ |
| $c_j$ | 0.644 | $-0.527$ | 0.971 | $-0.076$ | $-0.924$ | $-0.651$ |
| $z'(c_j)$ | 13 | $-11$ | $-20$ | $-2$ | 19 | 14 |
| $d_j$ | $-0.765$ | $-0.826$ | 0.319 | $-0.999$ | 0.221 | $-0.879$ |
| $z'(d_j)$ | $-16$ | $-17$ | 7 | 20 | 5 | $-18$ |

Table 5 shows $a_j, b_j, c_j, d_j, z'(a_j), z'(b_j), z'(c_j)$ and $z'(d)$. As the result, $X_j$'s $(1 \leq j \leq 6)$ are as follows:
$$X_1 \leftarrow \exp(i(-13x + 16y + 13z - 16w)),$$

$$X_2 \leftarrow \exp(i(6x + 19y - 11z - 17w)),$$
$$X_3 \leftarrow \exp(i(20x + 4y - 20z + 7w)),$$
$$X_4 \leftarrow \exp(i(15x - 16y - 2z + 20w)),$$
$$X_5 \leftarrow \exp(i(-4x - 19y + 19z + 5w)),$$
$$X_6 \leftarrow \exp(i(-19x - 4y + 14z - 18w)).$$

By calculating all the $\max\{|z'(A_k)|, |z'(B_k)|, |z'(C_k)|, |z'(D_k)|\}$'s, the integerized minimal maximum frequency $\min_{1 \leq k \leq 3^n - 1}\{\max\{|z'(A_k)|, |z'(B_k)|, |z'(C_k)|, |z'(D_k)|\}\}$ is obtained. The value is 5. The integerized maximal maximum frequency $\max_{1 \leq k \leq 3^n - 1}\{\max\{|z'(A_k)|, |z'(B_k)|, |z'(C_k)|, |z'(D_k)|\}\} = \sum_{i=1}^{6}|d_i| = 83$. Therefore, let $l=85$.

$$\frac{1}{85^4} \sum_{m_1=1}^{85} \sum_{m_2=1}^{85} \sum_{m_3=1}^{85} \sum_{m_4=1}^{85} f = -0.8125,$$

that is, $C$ is $-0.8125$. The process of the calculation is omitted. Theorem 11 says that the constant term $C$ is $\frac{1}{2^{3M}}(2k - 2^{3M})$. The number of satisfying assignments of $(X \vee Y \vee Z) \wedge (X \vee Y \vee \bar{Z})$ is 6. By substituting $k = 6$, and $M = 2$, $\frac{1}{2^{3M}}(2k - 2^{3M}) = -0.8125$. It has been verified that the experimental result equals Theorem 11.

In the above example, the minimal maximum frequency was obtained by calculating all the $\max\{|A_k|, |B_k|, |C_k|, |D_k|\}$'s. However, if the minimal maximum frequency decreases in polynomial order and the order is known, we can intergerize the frequencies without obtaining the minimal maximum frequency. For example, if the minimal maximum frequency decreases in the order of $n^{-3}$, we can integerize frequencies by multiplying the frequencies by $n^4$ (Details will be explained in Section 11) and rounding them to integers. The maximal maximum frequency is less than or equal to $n$. The integerized maximal maximum integer is less than or equal to $n \times n^4 = n^5$. Therefore, $l \cong n^5$.

## 10. Numerical experiments of the minimal maximum frequencies of four-variable complex trigonometric functions

### 10.1 Numerical experiment 1
Let us assign four-variable complex trigonometric functions to the variables in $f$ as follows:
$$X_j \leftarrow \exp(i(a_j x + b_j y + c_j z + d_j w)) \ (1 \leq j \leq n), \text{ where}$$
$$a_j = \sin(n^2 + j),$$
$$b_j = \sin\left((n^2 + j)\left(1 + \frac{\pi}{2(n^2 + 1)}\right)\right),$$
$$c_j = \sin\left((n^2 + j)\left(1 + \frac{3\pi}{n^2 + 1}\right)\right), \text{ and}$$
$$d_j = \sin\left((n^2 + j)\left(1 + \frac{3\pi}{n^2 + 1} + \frac{\pi}{2(n^2 + 1)}\right)\right),$$

that is, $u = n^2, p = 1, v = \frac{3\pi}{n^2 + 1}$ and $h = \frac{\pi}{2}\frac{1}{n^2 + 1}$.

Table 6 shows the minimal maximum frequencies. In Table 6, $m.m.f$ stands for the minimal maximum frequency. Figure 8 shows the logarithms of the minimal maximum frequencies. Figure 9 shows the logarithms of the minimal maximum frequencies of four-variable complex trigonometric functions(blue dots), the logarithms of the minimal maximum frequencies of two-variable complex trigonometric functions(yellow dots), and the logarithms of the minimal frequencies of one-variable complex trigonometric functions(green dots). As Figure 9 shows, four-variable complex trigonometric function is the best. Figure 10 shows the approximation by a logarithm function. Figure 11 shows the approximation by an irrational function. From Figure 10 and Figure 11, it is understood that the approximation by the logarithm function is better than that by the irrational function.

$$\log(m.m.f) = -3.31\log(0.4n) \to m.m.f = \frac{2.5^{3.31}}{n^{3.31}}$$

Therefore, the minimal maximum frequency may decrease in polynomial order.

Table 6  The minimal maximum frequencies (numerical experiment 1)

| $n$ | $m.m.f$ | $n$ | $m.m.f$ | $n$ | $m.m.f$ | $n$ | $m.m.f$ |
|---|---|---|---|---|---|---|---|
| 2 | 0.959 | 9 | 0.0214 | 16 | 0.00145 | 23 | 0.00103 |
| 3 | 0.839 | 10 | 0.0102 | 17 | 0.00108 | 24 | 0.00107 |
| 4 | 0.812 | 11 | 0.00704 | 18 | 0.000700 | 25 | 0.000555 |
| 5 | 0.465 | 12 | 0.00308 | 19 | 0.000933 | 26 | 0.000331 |
| 6 | 0.278 | 13 | 0.00271 | 20 | 0.00103 | 27 | 0.000255 |
| 7 | 0.0484 | 14 | 0.00185 | 21 | 0.00105 | 28 | |
| 8 | 0.0438 | 15 | 0.00175 | 22 | 0.000996 | 29 | |

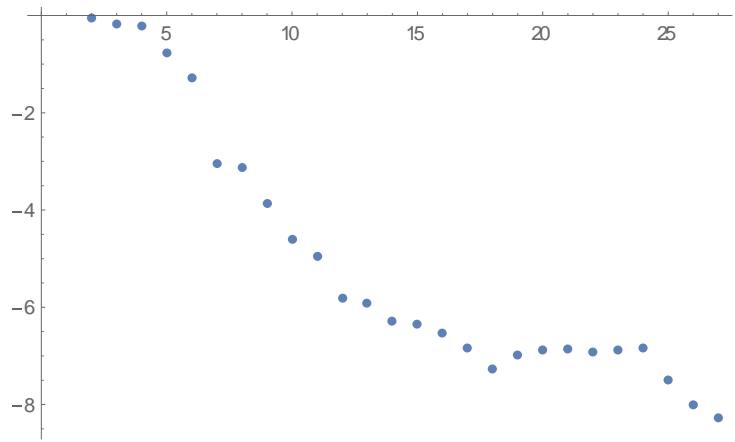

**Figure 8** The minimal maximum frequencies (logarithm)(numerical experiment 1)

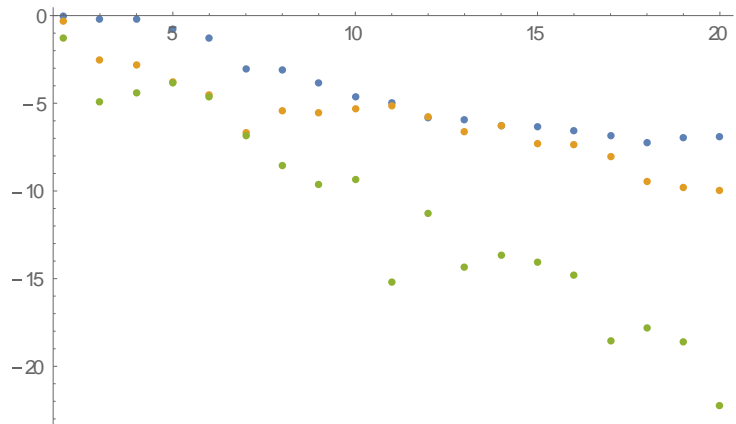

**Figure 9** The minimal maximum frequencies of four-variable complex trigonometric functions (logarithm), the minimal maximum frequencies of two-variable complex trigonometric functions (logarithm), and the minimal frequencies of one-variable complex trigonometric functions (logarithm)

### 10.2 Numerical experiment 2
Let
$$u = n^3, p = 1, v = \frac{3\pi}{n^3+1} \text{ and } h = \frac{\pi}{2}\frac{1}{n^3+1}.$$
$a_j, b_j, c_j$ and $d_j$ are as follows:

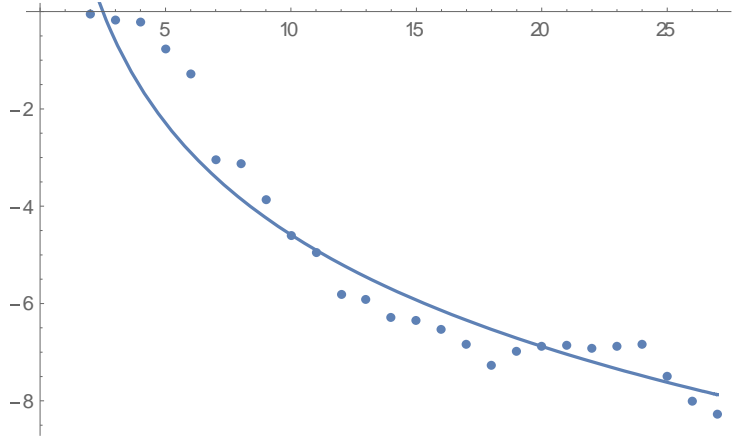

**Figure 10**  $-3.31\log(0.4n)$

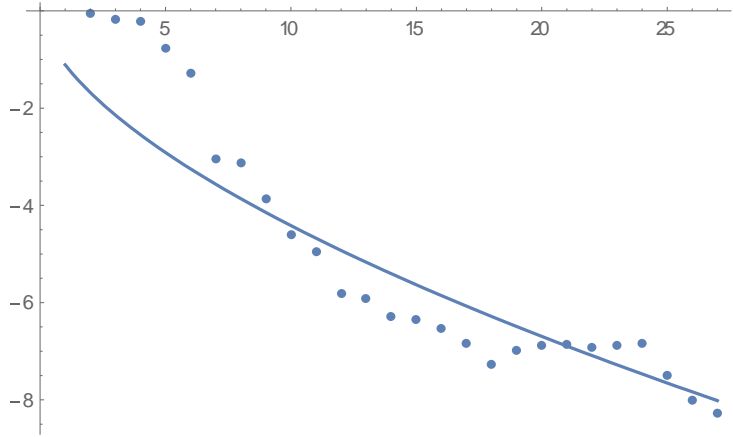

**Figure 11**  $-1.11n^{0.6}$

$$a_j = \sin(n^3 + j),$$
$$b_j = \sin\left((n^3 + j)\left(1 + \frac{\pi}{2(n^3 + 1)}\right)\right),$$
$$c_j = \sin\left((n^3 + j)\left(1 + \frac{3\pi}{n^3 + 1}\right)\right), \text{and}$$
$$d_j = \sin\left((n^3 + j)\left(1 + \frac{3\pi}{n^3 + 1} + \frac{\pi}{2(n^3 + 1)}\right)\right).$$

Table 7 shows the minimal maximum frequencies. Figure 12 shows the logarithms of the minimal maximum frequencies. Figure 13 shows the approximation by a logarithm function. Figure 14 shows the approximation by an irrational function. As Figure 13 and Figure 14 show, the approximation by the logarithm function is better than that by the irrational function.

$$\log(m.m.f) = -3.64\log(0.4n) \rightarrow m.m.f = \frac{2.5^{3.64}}{n^{3.64}}$$

Therefore, the minimal maximum frequency may decrease in polynomial order. The above two numerical experiments showed that the minimal maximum frequency may decrease in polynomial order. Therefore the computational complexity of the algorithm may be polynomial.

**Table 7** The minimal maximum frequencies (numerical experiment 2)

| $n$ | $m.m.f$ | $n$ | $m.m.f$ | $n$ | $m.m.f$ | $n$ | $m.m.f$ |
|---|---|---|---|---|---|---|---|
| 2 | 0.911 | 9 | 0.00844 | 16 | 0.00141 | 23 | 0.000318 |
| 3 | 0.468 | 10 | 0.00826 | 17 | 0.00131 | 24 | 0.000325 |
| 4 | 0.168 | 11 | 0.00693 | 18 | 0.000742 | 25 | 0.000247 |
| 5 | 0.0581 | 12 | 0.00303 | 19 | 0.000643 | 26 | 0.000194 |
| 6 | 0.00757 | 13 | 0.00240 | 20 | 0.000605 | 27 | 0.0000697 |
| 7 | 0.00911 | 14 | 0.00277 | 21 | 0.000570 | | |
| 8 | 0.00866 | 15 | 0.00262 | 22 | 0.000426 | | |

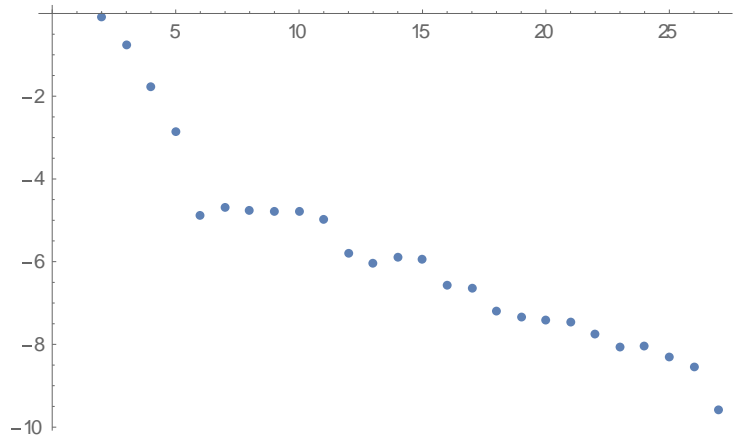

**Figure 12**  The minimal maximum frequencies（logarithm）(numerical experiment 2)

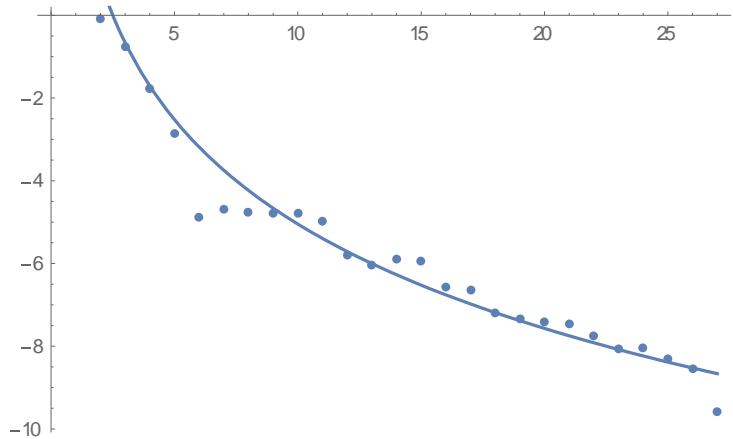

**Figure 13**  $-3.64\log 0.4n$

Figure 15 shows the logarithms of the minimal maximum frequencies of numerical experiment 1 and numerical experiment 2. Blue dots denote the minimal maximum frequencies of numerical experiment 1 and yellow dots denote the minimal maximum frequencies of numerical experiment 2. As Figure 15 shows, the result of numerical experiment 1 is better than that of numerical experiment 2.

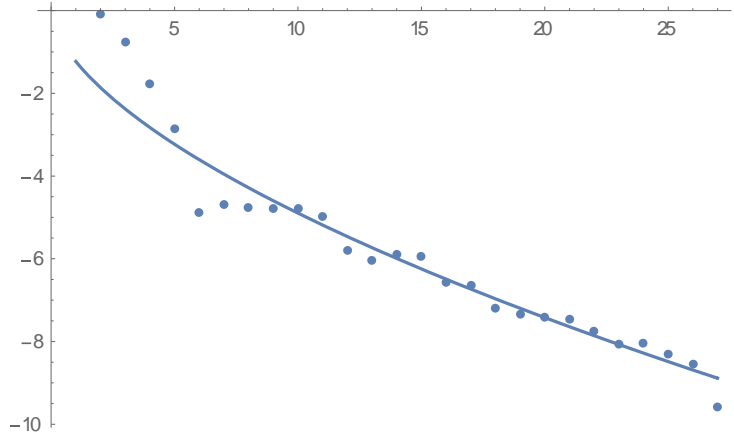

**Figure 14** $-1.23n^{0.6}$

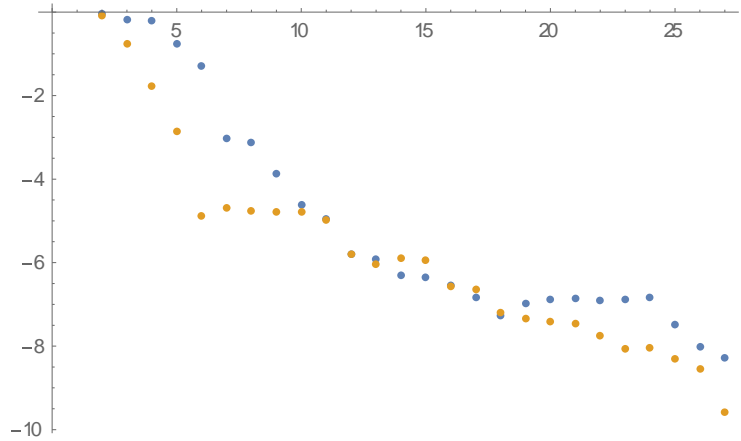

**Figure 15** Numerical experiment 1 and numerical experiment 2

## 11. Discussion

### 11.1 Integerization

For the integerization, frequencies are multiplied by a certain integer. How to determine the integer is explained. Assume that $m.m.f = \frac{p}{n^k}$, where $p$ is a positive real number, $n$ is the number of variables and $k$ is an integer.

**Theorem 19.** *For the integerization $z'(\cdot)$ of frequencies, the frequencies are multiplied by $n^l$, where $l$ is an integer such that $l \geq k + 1 - \log_n p$.*

*Proof.* Let frequencies be denoted by $a_i (1 \leq i \leq n)$. Then $m.m.f$ is described as follows:

$$m.m.f = \sum_{i=1}^{n} e_i a_i \ (e_i = -1, 0 \text{ or } 1),$$

and, since $m.m.f = \frac{p}{n^k}$, the following formula holds:

$$\sum_{i=1}^{n} e_i a_i = \frac{p}{n^k}.$$

Let the above formula be multiplied by $n^l$, the left-hand side is as follows:

$$n^l \sum_{i=1}^{n} e_i a_i = \sum_{i=1}^{n} e_i n^l a_i = \sum_{i=1}^{n} e_i (r(n^l a_i) + \alpha_i) = \sum_{i=1}^{n} e_i r(n^l a_i) + \sum_{i=1}^{n} e_i \alpha_i,$$

where $r(\cdot)$ stands for rounding and $\alpha_i$ is the fractional part. Therefore the following formula is obtained:

$$\sum_{i=1}^{n} e_i r(n^l a_i) + \sum_{i=1}^{n} e_i \alpha_i = n^{l-k} p.$$

Since

$$-n < \sum_{i=1}^{n} e_i \alpha_i < n,$$

$$n^{l-k} p - n < \sum_{i=1}^{n} e_i r(n^l a_i) < n^{l-k} p + n.$$

Since $\sum_{i=1}^{n} e_i r(n^l a_i)$ is the integerized minimal maximum frequency,

$$1 \leq \sum_{i=1}^{n} e_i r(n^l a_i).$$

Thus,

$$n^{l-k} p - n \geq 0 \rightarrow l \geq k + 1 - \log_n p. \qquad \square$$

**Theorem 20.** *For large $n$, $l \gtrsim k + 1$.*
*Proof.* For large $n$, $\log_n p \cong 0$. Therefore $l \gtrsim k + 1$. $\qquad \square$

**11.2 Computational complexity**
If the minimal maximum frequencies of two-variable complex trigonometric functions are approximated by $-4.4\log(0.5n)$ in Figure 4. Then $m.m.f = \frac{2^{4.4}}{n^{4.4}}$. For integerizing the minimal maximum frequency, let us multiply it by $n^{5.4}$(Theorem 20). The maximal maximum frequency is less than or equal to $n$. Therefore, the integerized maximal maximum frequency is less than or equal to $n^{6.4}$ and $l \cong n^{6.4}$. As the number of variables is two, the calculation for the sum $S_f$ in Theorem 17 needs $n^{6.4 \times 2} = n^{12.8} \cong n^{13}$ times of calculating $f$. As $n = 3M$, $n^{13} = (3M)^{13}$. However, this estimation may be optimistic.

Let us consider the results of four-variable complex trigonometric functions. As the result of numerical experiment 1 is better than that of numerical experiment 2, let us discuss the computational complexity based on the result of numerical experiment 1. In numerical experiment 1, $m.m.f = \frac{2.5^{3.31}}{n^{3.31}}$ in Figure 10. For integerizing the minimal maximum frequency, let us multiply it by $n^{4.31}$(Theorem 20). The maximal maximum frequency is less than or equal to $n$. Therefore, the integerized maximal maximum frequency is less than or equal to $n^{5.31}$ and $l \cong n^{5.31}$. As the number of variables is four, the calculation for the sum $S_f$ in Theorem 18 needs $n^{5.31 \times 4} = n^{21.24} \cong n^{21}$ times of calculating $f$. As $n = 3M$, $n^{21} = (3M)^{21}$.

There are $18M$ trigonometric functions in $f$. For simplification, let the computational complexity of trigonometric function be $O(n^2)$. (This is not the fastest evaluation.) Therefore the computational complexity of computing trigonometric functions in $f$ is $O(n^2 M)$. Since the computational complexity of four arithmetic operations in $f$ can be ignored, the computational complexity of $f$ is $O(n^2 M)$. The computational complexity of the algorithm is $O(n^2 M \times M^{21}) = O(M^{24})$.

If the minimal maximum frequency decreases in $\exp(-1.11 n^{0.6})$ in Figure 11, the computational complexity of the algorithm is subexponential.

The reason why the computational complexity of the algorithm is small lies in the facts shown in Table 3. In the case of one variable, the minimal frequency decreases in exponential order. On the other hand, in the case of two variables and in the case of four variables, the minimal maximum frequency decreases more slowly than that of one variable. The exponential decrease of the minimal frequency in the case of one variable can be regarded as a combinatorial explosion. In the case of two variables, the combinatorial explosion is suppressed owing to $\max\{|A_k|, |B_k|\}$. In the case of four variables, the combinatorial explosion is more suppressed owing to $\max\{|A_k|, |B_k|, |C_k|, |D_k|\}$.

**11.3 On parameters**
There are several parameters in the algorithm presented in this paper.

1. The number of the variables in complex trigonometric functions
   Two-variable complex trigonometric functions may work well. Four-variable complex trigonometric functions work well. The numbers greater than four probably work well. The author conducted the numerical experiment with six-variable complex trigonometric functions, the result of which is similar to that of four-variable complex trigonometric functions.
2. $p$
   In this paper, $p = 1$. Whether the algorithm works well or not largely depends on $p$. For example, $p = 0$ may be the worst. The logarithms of the minimal maximum frequencies with $p = 0$ are shown in Figure 16. The other parameters are the same as those of numerical experiment 1. Blue dots denote the minimal maximum frequencies with $p = 0$, and yellow dots denote the minimal maximum frequencies with $p = 1$(numerical experiment 1).

   The logarithms of the minimal maximum frequencies with $p = 2.5$ are shown in Figure 17. The other parameters are the same as those of numerical experiment 1. The minimal maximum frequencies with $p = 2.5$ may be approximated by a linear function, by an irrational function or by a logarithm function. As far as the author investigated, $p = 1$ is the best.

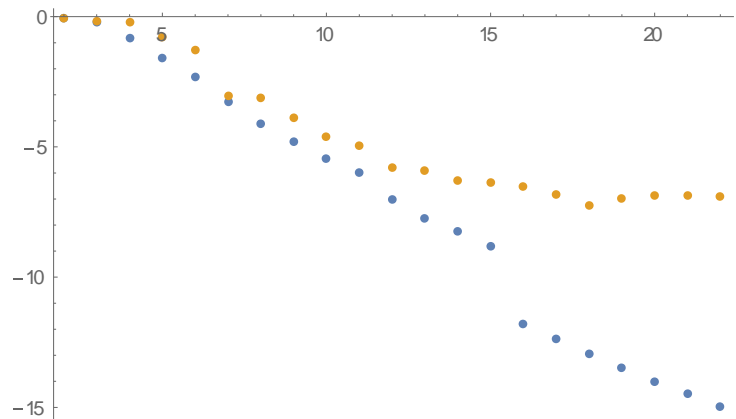
**Figure 16** The minimal maximum frequencies with $p = 0$

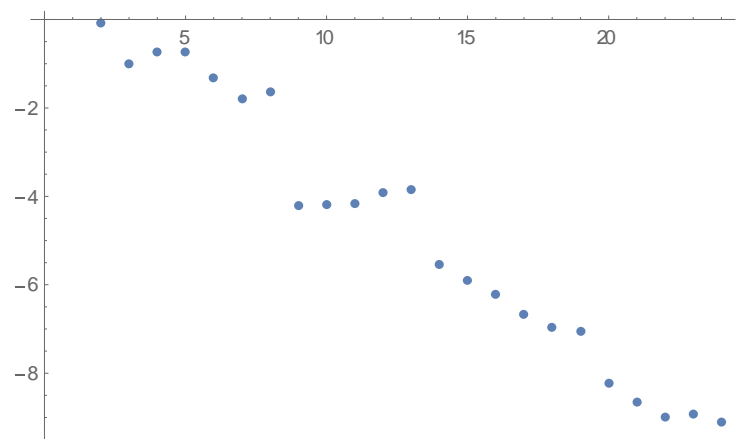
**Figure 17** The minimal maximum frequencies with $p = 2.5$

3. $u$

   $u = 2n$ does not work well. $u = n^2$ and $u = n^3$ work well. $u = n^2$ works better than $u = n^3$.

4. $v$

   In this paper, $v = \frac{3\pi}{n^2+1}$ or $\frac{3\pi}{n^3+1}$. Other values may also work well.

5. $h$

   In this paper, $h = \frac{\pi}{2}\frac{1}{n^2+1}$ or $\frac{\pi}{2}\frac{1}{n^3+1}$. Other values may also work well.

**11.4 Others**

1. In this paper, $a_j = \sin((u+j)p)$, $b_j = \sin((u+j)(p+h))$, $c_j = \sin((u+j)(p+v))$, and $d_j = \sin((u+j)(p+v+h))$. The author conducted the numerical experiments with

$$a_j = \sum_{k=1}^{n} q_{jk}\sin((u+k)p),$$

$$b_j = \sum_{k=1}^{n} r_{jk}\sin((u+k)(p+h)),$$

$$c_j = \sum_{k=1}^{n} s_{jk}\sin((u+k)(p+v)) \text{ and}$$

$$d_j = \sum_{k=1}^{n} t_{jk}\sin((u+k)(p+v+h)),$$

where $q_{jk}, r_{jk}, s_{jk}$ and $t_{jk}$ were determined by random variables. The result showed that the above $a_j, b_j, c_j$ and $d_j$ work better than $a_j, b_j, c_j$ and $d_j$ in this paper. The explanation is omitted, because it is complicated.

2. The author applied a few techniques (for example, moving average) to delete the constrictions in Figure 7, but they did not work well.

3. The numerical experiments of the minimal maximum frequency were conducted with $n \leq 27$. The numerical experiments need a lot of computing time. For example, the computing time of the numerical experiment with $n = 27$ by 51 parallel processings was about 10 days. The numerical experiments with $27 \leq n$ may be needed. However, the theoretical proof is more needed.

**12. Conclusions**

This paper has presented an algorithm for 3-SAT problems. The algorithm outputs the number of satisfying assignments. The computational complexity of the algorithm may be polynomial in the number of clauses. Future work includes the proof.

**Appendix**   The process of the expansion in Example 5

$$-\frac{23}{32} + \frac{3x_1}{32} + \frac{3x_2}{32} - \frac{3x_1x_2}{32} + \frac{3x_3}{32} - \frac{x_1x_3}{4} - \frac{3}{32}x_1^2x_3 + \frac{x_2x_3}{32} - \frac{1}{8}x_1x_2x_3 + \frac{3}{32}x_1^2x_2x_3 - \frac{3}{32}x_1x_3^2$$

$-\frac{1}{32}x_1^2x_3^2 - \frac{1}{32}x_1x_2x_3^2 + \frac{1}{32}x_1^2x_2x_3^2 + \frac{3x_4}{32} + \frac{x_1x_4}{32} + \frac{5x_2x_4}{16} + \frac{1}{16}x_1x_2x_4 + \frac{3}{32}x_2^2x_4 - \frac{3}{32}x_1x_2^2x_4$
$-\frac{3x_3x_4}{32} - \frac{1}{8}x_1x_3x_4 - \frac{1}{32}x_1^2x_3x_4 + \frac{1}{16}x_2x_3x_4 - \frac{1}{4}x_1x_2x_3x_4 - \frac{1}{16}x_1^2x_2x_3x_4 + \frac{1}{32}x_2^2x_3x_4$
$-\frac{1}{8}x_1x_2^2x_3x_4 + \frac{3}{32}x_1^2x_2^2x_3x_4 + \frac{3}{32}x_1x_3^2x_4 + \frac{1}{32}x_1^2x_3^2x_4 - \frac{1}{16}x_1x_2x_3^2x_4 - \frac{1}{16}x_1^2x_2x_3^2x_4$
$-\frac{1}{32}x_1x_2^2x_3^2x_4 + \frac{1}{32}x_1^2x_2^2x_3^2x_4 + \frac{3}{32}x_2x_4^2 + \frac{1}{32}x_1x_2x_4^2 + \frac{1}{32}x_2^2x_4^2 - \frac{1}{32}x_1x_2^2x_4^2 - \frac{3}{32}x_2x_3x_4^2$
$-\frac{1}{8}x_1x_2x_3x_4^2 - \frac{1}{32}x_1^2x_2x_3x_4^2 - \frac{1}{32}x_2^2x_3x_4^2 + \frac{1}{32}x_1^2x_2^2x_3x_4^2 + \frac{3}{32}x_1x_2x_3^2x_4^2 + \frac{1}{32}x_1^2x_2x_3^2x_4^2$
$+\frac{1}{32}x_1x_2^2x_3^2x_4^2 - \frac{1}{32}x_1^2x_2^2x_3^2x_4^2$

$x_1{}^2 \to 1$ is applied.
$-\frac{23}{32} + \frac{3x_1}{32} + \frac{3x_2}{32} - \frac{3x_1x_2}{32} - \frac{x_1x_3}{4} + \frac{x_2x_3}{8} - \frac{1}{8}x_1x_2x_3 - \frac{x_3^2}{32} - \frac{3}{32}x_1x_3^2 + \frac{1}{32}x_2x_3^2 - \frac{1}{32}x_1x_2x_3^2$
$+\frac{3x_4}{32} + \frac{x_1x_4}{32} + \frac{5x_2x_4}{16} + \frac{1}{16}x_1x_2x_4 + \frac{3}{32}x_2^2x_4 - \frac{3}{32}x_1x_2^2x_4 - \frac{x_3x_4}{8} - \frac{1}{8}x_1x_3x_4 - \frac{1}{4}x_1x_2x_3x_4$
$+\frac{1}{8}x_2^2x_3x_4 - \frac{1}{8}x_1x_2^2x_3x_4 + \frac{1}{32}x_3^2x_4 + \frac{3}{32}x_1x_3^2x_4 - \frac{1}{16}x_2x_3^2x_4 - \frac{1}{16}x_1x_2x_3^2x_4 + \frac{1}{32}x_2^2x_3^2x_4$
$-\frac{1}{32}x_1x_2^2x_3^2x_4 + \frac{3}{32}x_2x_4^2 + \frac{1}{32}x_1x_2x_4^2 + \frac{1}{32}x_2^2x_4^2 - \frac{1}{32}x_1x_2^2x_4^2 - \frac{1}{8}x_2x_3x_4^2 - \frac{1}{8}x_1x_2x_3x_4^2$
$+\frac{1}{32}x_2x_3^2x_4^2 + \frac{3}{32}x_1x_2x_3^2x_4^2 - \frac{1}{32}x_2^2x_3^2x_4^2 + \frac{1}{32}x_1x_2^2x_3^2x_4^2$

$x_2{}^2 \to 1$ is applied.
$-\frac{23}{32} + \frac{3x_1}{32} + \frac{3x_2}{32} - \frac{3x_1x_2}{32} - \frac{x_1x_3}{4} + \frac{x_2x_3}{8} - \frac{1}{8}x_1x_2x_3 - \frac{x_3^2}{32} - \frac{3}{32}x_1x_3^2 + \frac{1}{32}x_2x_3^2 - \frac{1}{32}x_1x_2x_3^2$
$+\frac{3x_4}{16} - \frac{x_1x_4}{16} + \frac{5x_2x_4}{16} + \frac{1}{16}x_1x_2x_4 - \frac{1}{4}x_1x_3x_4 - \frac{1}{4}x_1x_2x_3x_4 + \frac{1}{16}x_3^2x_4 + \frac{1}{16}x_1x_3^2x_4$
$-\frac{1}{16}x_2x_3^2x_4 - \frac{1}{16}x_1x_2x_3^2x_4 + \frac{x_4^2}{32} - \frac{1}{32}x_1x_4^2 + \frac{3}{32}x_2x_4^2 + \frac{1}{32}x_1x_2x_4^2 - \frac{1}{8}x_2x_3x_4^2 - \frac{1}{8}x_1x_2x_3x_4^2$
$-\frac{1}{32}x_3^2x_4^2 + \frac{1}{32}x_1x_3^2x_4^2 + \frac{1}{32}x_2x_3^2x_4^2 + \frac{3}{32}x_1x_2x_3^2x_4^2$

$x_3{}^2 \to 1$ is applied.
$-\frac{3}{4} + \frac{x_2}{8} - \frac{x_1x_2}{8} - \frac{x_1x_3}{4} + \frac{x_2x_3}{8} - \frac{1}{8}x_1x_2x_3 + \frac{x_4}{4} + \frac{x_2x_4}{4} - \frac{1}{4}x_1x_3x_4 - \frac{1}{4}x_1x_2x_3x_4 + \frac{1}{8}x_2x_4^2$
$+\frac{1}{8}x_1x_2x_4^2 - \frac{1}{8}x_2x_3x_4^2 - \frac{1}{8}x_1x_2x_3x_4^2$

$x_4{}^2 \to 1$ is applied.

$-\frac{3}{4} + \frac{x_2}{4} - \frac{x_1x_3}{4} - \frac{1}{4}x_1x_2x_3 + \frac{x_4}{4} + \frac{x_2x_4}{4} - \frac{1}{4}x_1x_3x_4 - \frac{1}{4}x_1x_2x_3x_4$